%% file: Schaefer_etal_CRs_MoleculeRelease_SphericalMatrixCarrier.tex
\documentclass[journal, 12pt, onecolumn, draftclsnofoot]{IEEEtran}

\usepackage[T1]{fontenc}% optional T1 font encoding

\IEEEoverridecommandlockouts
\usepackage{cite}
\usepackage{amsmath,amssymb,amsfonts}
\usepackage{bm}
\usepackage{nicefrac}
\usepackage{siunitx}
\usepackage{graphicx}
\usepackage{textcomp}
\usepackage{xcolor}
\usepackage{subcaption}
\usepackage[ruled,linesnumbered]{algorithm2e} % vlined
\SetAlFnt{\small}
\SetAlCapFnt{\small}
\SetAlCapNameFnt{\small}
\def\BibTeX{{\rm B\kern-.05em{\sc i\kern-.025em b}\kern-.08em
    T\kern-.1667em\lower.7ex\hbox{E}\kern-.125emX}}

\usepackage{cite}
\usepackage{subcaption}
    
\usepackage{tikz}
\usetikzlibrary{math}
\usetikzlibrary{shapes.geometric,calc}
\usetikzlibrary{arrows.meta, quotes, angles}
\usetikzlibrary{decorations}
\usetikzlibrary{chains,shapes.misc,positioning,scopes}
\usepackage{circuitikz}
\usetikzlibrary{circuits.ee.IEC}
\usetikzlibrary{decorations}
\usetikzlibrary{dsp}
\usetikzlibrary{shadows}
\usepackage{subcaption}

\newcommand{\acs}{\frac{A}{C_\mathrm{s}}}
\newcommand{\nacs}{\nicefrac{A}{C_\mathrm{s}}}
\newcommand{\csa}{\frac{C_\mathrm{s}}{A}}
\newcommand{\trel}{t_\mathrm{rel}}
\newcommand{\tabs}{t_\mathrm{abs}}
\newcommand{\tmax}{t_\mathrm{max}}
\newcommand{\tran}{^{\scriptsize{\mathrm{T}}}}

\newcommand{\rrx}{r_\mathrm{\tiny RX}}

\newcommand{\Dm}{D_\mathrm{m}}
\newcommand{\Dc}{D_\mathrm{c}}

\newcommand{\ratio}{\tau}

\newcommand{\erf}[1]{\mathrm{erf}\!\left(#1\right)}
\newcommand{\erfc}[1]{\mathrm{erfc}\!\left(#1\right)}
\newcommand{\expI}[1]{\mathrm{exp}\!\left(#1\right)}
\newcommand{\erfcInv}[1]{\mathrm{erfcinv}\!\left(#1\right)}

\newcommand{\dint}[1]{\,\mathrm{d}#1}
\definecolor{mygreen}{RGB}{80, 220, 100}

\begin{document}

% End-to-end Channel Model for Controlled-Release Drug Delivery using Practical Drug Carriers
% 
%\title{Channel Responses for the Molecule Release from Polymer based Drug Carrier Matrices}

\title{Channel Responses for the Molecule Release from Spherical Homogeneous Matrix Carriers}

\author{Maximilian Sch\"afer,
        Yolanda Salinas, Alexander Ruderer, Franz Enzenhofer, Oliver Br\"uggemann,     
Ram\'{o}n Mart\'{i}nez-M\'{a}\~{n}ez, Rudolf Rabenstein, Robert~Schober, 
and~Werner Haselmayr% <-this
%\author{Maximilian Sch\"afer,~\IEEEmembership{Member,~IEEE,}
%        Yolanda Salinas, Alexander Ruderer, Franz Enzenhofer, Oliver Br\"uggemann,     
%Robert~Schober,~\IEEEmembership{Fellow,~IEEE}
%and~Werner Haselmayr,~\IEEEmembership{Member,~IEEE,}% <-this % stops a space
%\thanks{This work was supported by the German Research Foundation (DFG) under grant number SCHO 831/14-1, }%
%\thanks{M. Sch\"afer, A. Ruderer and R. Schober are with the Institute for Digital Communications, Friedrich-Alexander-Universität Erlangen-Nürnberg, 91058 Erlangen, Germany (e-mail: max.schaefer@fau.de; robert.schober@fau.de). 
%Y. Salinas, F. Enzenhofer, O. Br\"uggemann and W. Haselmayr are with the Chair of Multimedia Communications and Signal Processing, Friedrich-Alexander-Universität Erlangen-Nürnberg, 91058 Erlangen, Germany (e-mail: rudolf.rabenstein@fau.de).
%}% <-this % stops a space
\thanks{This manuscript has been submitted in part for presentation at the IEEE Global Communications Conference, 2021 \cite{schaefer:globecome:2021}.}
\vspace*{-7ex}
}

\maketitle

% ===================================================================
\input{abstract}

% ===================================================================
\section{Introduction}
\label{sec:intro}
\input{introduction}

%===========================================================
\section{Spherical Matrix Systems} %Drug Release of Homogeneous Matrix Systems
\label{sec:math}
\input{math}

%===================================================================
\section{Modeling the Release from Matrix Systems} 
\label{sec:drug}
\input{drug_release}

%===================================================================
%\section{Relation to Realistic Release Processes} 
%\label{sec:param}
%\input{real_param}

%===========================================================
\section{Channel Response for Spherical Matrix}
\label{sec:channel}

\input{channel}

%===========================================================
\section{Approximate Channel Responses for Limiting Regimes}
\label{sec:regimes}
\input{regimes}

%===========================================================
\section{Numerical Results}
\label{sec:eval}
\input{num_eval}

% ===================================================================
\section{Conclusions}
\label{sec:concl}
\input{conclusions}

\appendices
\section{Upper Bound for the Approximation Error in the Channel Dominated Regime}
\label{sec:ap_up_chDom}
\input{ap_ub_chDom}

\section{Upper Bound for the Approximation Error in the Release Dominated Regime}
\label{sec:ap_up_reDom}
\input{ap_ub_reDom}

\bibliographystyle{IEEEtran}
{\footnotesize
	\bibliography{./IEEEabrv,./msnp}}

\end{document}

%% file: abstract.tex
\begin{abstract}
Molecular communications is a promising framework for the design of controlled-release drug delivery systems. 
Under this framework, drug carriers, diseased cells, and the channel in between are modeled as transmitters, absorbing receivers, and diffusive channel, respectively. 
However, existing works on drug delivery systems consider only simple drug carrier models, which limits their practical applicability. 
In this paper, we investigate diffusion-based spherical matrix-type drug carriers, which are employed in medical applications. 
In a matrix carrier, the drug molecules are dispersed in the matrix core and diffuse from the inner to the outer layers of the carrier once immersed in a dissolution medium. 
We derive the channel response of the matrix carrier transmitter for an absorbing receiver. The results are validated by particle-based simulations and compared  with commonly used point and transparent spherical transmitters to highlight the necessity of considering practical models. 
Moreover, we show that a transparent spherical transmitter, with the drug molecules uniformly distributed over the entire volume, is a special case of the considered matrix system. 
For this case, we provide an analytical expression for the channel response. 
Furthermore, we derive a criterion for evaluating whether the release process or the channel dynamics are more important for the characteristics of the channel response of a drug delivery system. 
For the limiting regimes, where only the release process or only the channel determine the behavior of the end-to-end system, we propose closed-form approximations for the channel response. 
Finally, as a scenario of practical relevance, we investigate the channel responses for the release of common therapeutic drugs, e.g., doxorubicin, from a diblock copolymer micelle acting as drug carrier. 
\end{abstract}

%\begin{IEEEkeywords}
%do we need key words
%\end{IEEEkeywords}

%% file: introduction.tex
Molecular communications (MC) considers the transmission of information using biochemical signals over multiple scales~\cite{Nakano_13}. 
Over the past few years, the MC paradigm has been exploited to gain more insight into the operation of biological systems, to control the behavior of such systems, and for the design and implementation of synthetic MC systems in the macro- and micro-/nanoscale~\cite{Farsad_16}. 
Human-made MC systems are expected to have applications in biomedical, environmental, and industrial engineering~\cite{Nakano_13, Farsad_16}. % BOOK: Chude_Okonkwo_19
%Widely found in nature  as it is essential for all living entities to retain their functionalities, for example the endocrine system releases signaling molecules (i.e., hormones) into the blood stream to communicate with distant target cells in order to regulate body activities. 

The MC framework is a promising approach to design drug delivery (DD) systems, where nanoparticle carriers deliver drug molecules to diseased cell sites and release them at the right time and rate~\cite{Chude_Okonkwo_17,Chahibi_17,Sutradhar2016,salinas_2020}. 
This targeted delivery reduces potential side effects on non-target sites and helps mitigate toxicity, drug wastage, and healthcare costs compared to conventional treatment options. 
Based on the MC paradigm, drug carriers, diseased cells, and the drug propagation are modeled as transmitters (TXs), receivers~(RXs), and random channel, respectively~\cite{Chude_Okonkwo_17,Chahibi_17}.   
The MC related research on DD systems can be categorized into three areas: \textit{i) target detection} aims to develop methods for localization of diseased cells and moving the drug carriers towards them~\cite{Nakano_17}; \textit{ii) drug propagation in the circulatory system} aims to develop models for the distribution of drug molecules or  carriers over time for the optimization of drug injection~\cite{Chahibi_13}; \textit{iii) controlled local drug release} aims to design an optimum controlled-release profile, assuming that the drug carriers are already near the diseased cells~\cite{Femminella_15,Cao_19,Salehi_18,Salehi_17,Lin_21}. 

In this work, we focus on \textit{controlled local drug release}. 
%Referring to the previous interpretation of a TDD system under MC paradigm, 
In particular, we focus on the analysis and design of TX models which provide a controlled local release of signaling (drug) molecules. 
Existing works in the MC literature study various aspects 
of DD systems. 
The mobility of multiple drug carriers due to diffusion and its influence on the amount of absorbed molecules is investigated in~\cite{Cao_19}. 
The impact of limited drug reservoir capacity~\cite{Salehi_18} and drug release rate optimization \cite{Salehi_17,Lin_21} have also been investigated for DD systems with multiple TXs. 
While these studies characterize and optimize the process of delivery under different conditions, they mostly rely on simple point TX models. 
However, the investigation of practical TX models for drug carriers is very important for the envisioned design of efficient DD systems. 
The most common TX models in the MC literature are point or simple spherical TX models which model an uncontrolled instantaneous release of molecules into the environment~\cite{noel:16}. 
Nevertheless, there are a few studies that consider more complex TX models \cite{Arjmandi_2016, schaefer:icc:2020, Huang2020}. 
In \cite{Arjmandi_2016}, the release of molecules from a biologically inspired TX is controlled by the opening and closing of ion-channels. 
In \cite{schaefer:icc:2020}, a spherical TX is considered with a spatially and temporally adjustable semi-permeable membrane to control the release of molecules. 
Recently, the authors in \cite{Huang2020} considered a membrane-fusion based TX, where the release of molecules encapsulated in a vesicle is based on the fusion of the vesicle and the TX membrane.

The objective of this work is to introduce and assess models for realistic drug carriers that are applied in practical pharmaceutical and chemical studies.
While the TX models in \cite{Arjmandi_2016, schaefer:icc:2020, Huang2020} may be considered as examples for reservoir based carriers, where the drug release is controlled by membrane functionalization, we consider spherical matrix type drug carriers, where drug molecules are dispersed inside the polymer core of the carrier 
%, which mostly consist of a polymer structure 
\cite{Macha_20, Chude_Okonkwo_19, Yanlong:pharmaceutics:2020}.
%\footnote{Polymers are mostly used as material due to their versatile properties~\cite{Macha_20}.} 
The release of drug molecules from polymer matrices is influenced by the physical and chemical properties of both, the polymer and the drug molecules. 
Prevalent release types are diffusion controlled, swelling controlled, and erosion controlled release \cite{Arifin2006, Siepmann2008}. 
The polymer matrix can be either compact or porous and in all three release mechanisms, the diffusion of drug molecules is involved. 
Practical examples of polymer matrices that are applied in DD systems include diblock copolymer micelles \cite{Sutton2007, Ahmad2014}, microspheres \cite{Macha_20,Arifin2006}, and nanogels \cite{Neamtu2017, Salinas2018}.
In this paper, we focus on the \textit{diffusion controlled release} from a non-biodegradable (non-erodible) polymer matrix system. 
Furthermore, we assume a compact polymer matrix yielding a homogeneous release of drug molecules.
In these systems, undissolved drug molecules are homogeneously distributed inside the polymer structure of the matrix. 
Once the matrix is immersed in a solution, the drug molecules start to dissolve and diffuse unconfined from the inner to the outer layers before propagating further into the surrounding medium. 
Due to the smaller amount of molecules located in the inner layers of the matrix and the increased diffusion distance, the release rate decreases \mbox{over time}. 
%
%Since the diffusion distance increases for molecules located further inside the carrier, 
%the release rate decreases over time.

There are various models for the release of drug molecules from homogeneous and porous matrix systems, see \cite{Siepmann2008} and \cite{Lee_11} for comprehensive overviews. 
However, to the best of the authors' knowledge, there is no general mathematical theory that can be applied to all types of drug release processes, as depending on the type of system, different physical, biological, and chemical processes occur during the drug release \cite{Siepmann2008}. 
In most practical studies heuristic approaches are employed, where a model for the drug release with several degrees of freedom is fitted to measurement data. A comprehensive overview on heuristic models is provided in \cite{Siepmann2008}. 
Although these models are very flexible, it is cumbersome to relate the estimated model parameters to the parameters of a particular drug carrier. 
Another approach is the modeling of the actual physical phenomena behind the release process. 
For example, the process of drug dissolution and matrix erosion can be described as a moving boundary diffusion problem. 
The first such model was presented by Higuchi in 1963 \cite{Higuchi:1963} and several similar models were proposed over the past decades \cite{Lee_11,Frenning:2004,Koizumi:1995}. 
Compared to the heuristic approach, these models provide a direct connection between the physical parameters of the carrier, the drug molecules and the amount of released drugs. 
%However, the necessity of too many system-specific parameters  sometimes narrows the applicability of these models to practical systems as the availability of all parameters is not guaranteed. 

All previously mentioned models do not consider the propagation of the released molecules in the surrounding environment and their absorption at an RX. 
However, these considerations are crucial for the investigation of DD systems based on the MC paradigm, and would allow for an optimization of the release process taking into account the channel and RX.
In this paper, we investigate a diffusive MC system employing a homogeneous matrix system as TX and an absorbing~RX, e.g., a~diseased cell or its nucleus. 
The obtained results provide the basis for the design of practical controlled-release DD systems. 
Our main contributions can be summarized as follows:
\vspace*{-1ex}
\begin{itemize}
  \item We discuss and analyze existing models for the release of molecules from spherical homogeneous matrix systems and derive an expression for the channel response (CR) for a matrix TX and an absorbing RX in a three-dimensional~(3D) unbounded environment. 
  \item For the special case where molecules are instantaneously released from the matrix TX, we derive an analytical expression for the CR. 
  This result corresponds to the CR of a transparent spherical TX, which is discussed in~\cite{noel:16} without providing a closed-form solution.
  \item We develop a particle-based simulation (PBS) model for the matrix TX to verify our theoretical results. Moreover, we compare the CR and the absorption rate of the matrix~TX with those of point and transparent spherical~TXs.
  \item We define two limiting regimes for the DD process, where either the release process or the channel dynamic determine the characteristic of the overall CR. For these limiting regimes, where only one of these processes is relevant for the behavior of the end-to-end system, we propose closed-form expressions to approximate the actual CR.
	\item We apply the proposed models for drug release to a realistic system. In particular, the release of two common therapeutic drugs from a diblock copolymer micelle is investigated. 
%	The CR between the realistic TX and an absorbing RX is analyzed.  
\end{itemize}

This paper extends the preliminary results in the conference version \cite{schaefer:globecome:2021} as follows:
We define two limiting regimes for the DD process, where either the release process or the channel dynamic determines the characteristic of the CR. 
For these limiting regimes, we derive closed-form expressions to approximate the CR which was obtained numerically in \cite{schaefer:globecome:2021}.
Second, we investigate the release process of therapeutic drugs from a diblock copolymer micelle as a practical drug carrier. 
Based on this analysis, we introduce further simplifications to the molecule release model presented in \cite{schaefer:globecome:2021}. 
Furthermore, we show that the proposed approximate closed-form CR is suitable for the characterization of DD from practical drug carriers.
%

% TODO update the remainder in the end!!! 
The remainder of this paper is organized as follows: 
In Section~\ref{sec:math}, we provide a mathematical description for the release of drug molecules from spherical homogeneous matrix systems and investigate the release of therapeutic drugs from a diblock copolymer micelles.
In Section~\ref{sec:drug}, we present models for the release of drug molecules. 
%which are related to the drug release from a polymer micelle carrier in Section~\ref{sec:param}.
We derive the CR of a matrix TX for an absorbing RX in Section~\ref{sec:channel}. 
In Section~\ref{sec:regimes}, we classify the transport of drug molecules from TX to RX into different regimes, where either the release process or the channel dynamics determine the characteristics of the CR. 
Numerical evaluations of the CR of a matrix TX and a validation through PBS are presented in Section~\ref{sec:eval}. 
%Moreover, comparisons with a point TX and a virtual sphere TX are provided. 
%and we present the CR for drug release from a practical carrier. 
Finally, Section~\ref{sec:concl} concludes the paper.

%% file: math.tex
%In this section, we investigate spherical matrix systems as a description of practical drug carriers. 
In this section, we discuss the most common drug release mechanisms from polymeric nanoparticles which are practical examples of spherical matrix systems. 
Then, we present a mathematical description for the release of molecules from spherical matrix systems.
Finally, we investigate the drug release from a diblock copolymer micelle, a practical drug carrier.

\subsection{Drug Release Mechanisms}
Polymeric nanoparticles are commonly employed as practical drug carriers \cite{Leong1988, Macha_20, Sutton2007, Ahmad2014} and due to their manufactured structure, they can be described as matrix systems \cite{Arifin2006}. 
The release of drug molecules from such polymer matrices depends on the physical and chemical characteristics of the polymer and drug molecules. 
Depending on the type of the drug carrier, there are three characteristic drug release mechanisms \cite{Arifin2006, Siepmann2008, Leong1988}: 
\begin{itemize}
	\item[(i)] \textit{Diffusion controlled release}: The molecule release is driven by diffusion of dissolved drug molecules from a non-biodegradable  matrix, e.g., a diblock copolymer micelle~\cite{Sutton2007, Ahmad2014},
	\item[(ii)] \textit{Swelling controlled release}: The molecule release is influenced by the polymer swelling that enhances the drug diffusion, e.g., in nanogels \cite{Neamtu2017, Salinas2018}, 
	\item[(iii)] \textit{Erosion controlled release}: The molecule release from biodegradable polymer matrices is controlled by matrix erosion due to the hydrolytic cleavage of polymer chains \cite{Lee_11, Arifin2006}. However, diffusion may be dominant when erosion is slow. 
\end{itemize}
Further effects that influence the drug release are the porosity of the matrix \cite{Lee_11}, and chemical interactions that may occur between the drug molecules and the polymer \cite{Sutton2007}. 
%
%While heuristic models are mostly employed in practical studies to handle the numerous occurring effects, they provide 
%
In this paper, we focus on \textit{diffusion controlled} release from non-biodegradable (non-erodible) matrix systems. 
%Furthermore, we assume that the polymer matrix is not porous, yielding a homogeneous release of drug molecules. 
Furthermore, we assume a compact polymer matrix, i.e., the propagation of dissolved drug molecules inside the matrix is not confined, yielding a homogeneous release of drug molecules. 
A practical example for this type of matrix system are, e.g., diblock copolymer micelles that are described further in Section~\ref{subsec:release_micelle}.
Depending on the type of drug molecules, chemical and physical interactions between the drug molecules and the matrix polymer may occur during the release process \cite{Sutton2007}. 
As we focus on diffusion controlled release, these effects are not considered in this paper and are left as an interesting topic for further work.
%
%In order to keep the number of occurring physical and chemical effects manageable, we neglect any chemical and physical interaction between the drug molecules and the matrix polymer.

\subsection{Mathematical Description of Drug Release from Homogeneous Matrix Systems}
\label{sec:drug_cons}
\begin{figure*}
	\centering
	\begin{subfigure}[b]{0.49\linewidth}
    \centering
            	\includegraphics[width=0.75\linewidth]{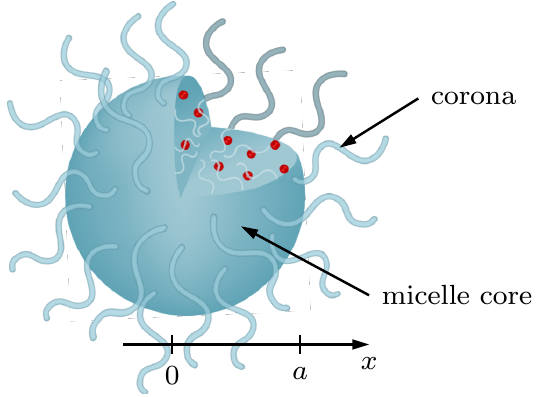}
            	\vspace*{-1ex}
            \caption{}
    \label{fig:micelle}
    \end{subfigure}
    \begin{subfigure}[b]{0.49\linewidth}
            	\centering
            	\includegraphics[width=0.9\linewidth]{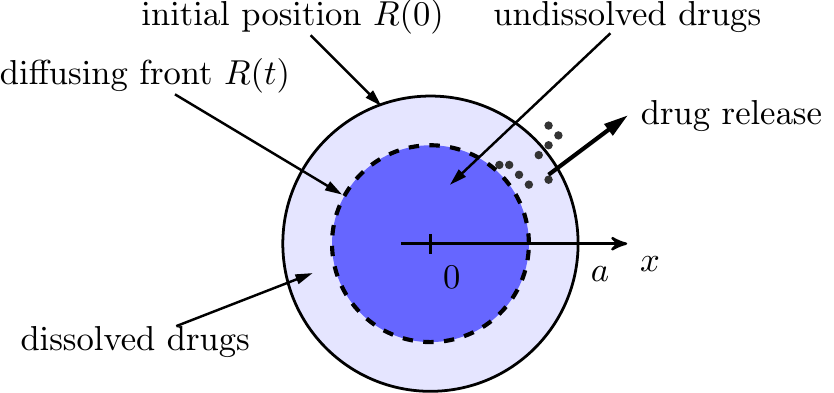}
            	\vspace*{-1ex}
            \caption{}
    \label{fig:1}
    \end{subfigure}
    \vspace{-3ex}
    \caption{\small (a) Schematic of a PEG-$b$-PLA diblock copolymer micelle, where the micelle core consists of poly(DL-lactide) (PLA) and is surrounded by the corona consisting of poly(ethylene glycol) (PEG). The drug molecules (red circles) are $\beta$-lapachone or doxorubicin molecules \cite{Sutton2007}. (b) Schematic release of a dispersed drug from a spherical non-erodible homogeneous matrix system as model for the drug release from the micelle core in (a).}
    \label{fig:_1}
    \vspace*{-5ex}
\end{figure*}

%\vspace*{-0.5ex}
%Fig.~\ref{fig:nanogel} shows an illustration of a polymeric nanogel loaded with drug molecules, a practical drug carrier \cite{referenz to be added}. 
%Initially, the drug molecules are dispersed in the nanogel. When the carrier is placed in a solution, the drug molecules are dissolved from the polymeric structure and diffuse inside the nanogel until they are finally released into the surrounding medium. 
%
%In the following, we present the mathematical description of the release process of molecules from a   of this process. 
Fig.~\ref{fig:micelle} illustrates a diblock copolymer micelle, a practical drug carrier (see Section~\ref{subsec:release_micelle}), and Fig.~\ref{fig:1} shows a two-dimensional schematic of the molecule release process from a \textit{spherical homogeneous non-erodible matrix system} as a model for the molecule release from the micelle core in Fig.~\ref{fig:micelle}.
%
%Fig.~\ref{fig:1} shows a two-dimensional schematic of the molecule release process from a \textit{spherical homogeneous non-erodible matrix system}, and Fig.~\ref{fig:micelle} illustrates a diblock copolymer micelle, a practical drug carrier (see Section~\ref{subsec:release_micelle}).
%
%
The matrix of radius $a$ is initially loaded with undissolved molecules (dark blue).
%, and the initial loading is denoted by $A$. 
The release process in a solution medium 
%with drug solubility $C_\mathrm{s}$ 
is modeled by a diffusing front $R(t)$ that defines the time dependent dissolution of molecules by the solution migrating in the matrix \cite{Higuchi:1963}. Dissolved molecules diffuse inside the matrix until they enter the surrounding medium at $x = a$. 
For the modeling of the release process, we make the following assumptions:
\begin{itemize}
	\item[A1)] Due to the \textit{homogeneous release} from the spherical matrix into an unbounded environment, the 3D system can be reduced to a one-dimensional system as both angular components can be neglected \cite{Lee_80,Higuchi:1963}.
	\item[A2)] Since the molecule carrying matrix is \textit{non-erodible}, the impact of an additional inward moving eroding front~(see~\cite[Fig.~1]{Lee_80}) is negligible. 
	\item[A3)] The moving boundary $R(t)$ (dashed line in Fig.~\ref{fig:1}) separates undissolved dispersed molecules (dark blue) and dissolved molecules (light blue).
\end{itemize}
The molecule release rate from the matrix mainly depends on the dimensionless ratio $\nacs$, with initial loading per unit volume $A$ and solubility of the drug molecules in the surrounding medium $C_\mathrm{s}$ \cite{Higuchi:1963}. 
%For example, a
Assuming a given initial loading $A$, molecules dissolve slowly for small values of solubility $C_\mathrm{s}$ (high $\nicefrac{A}{C_\mathrm{s}}$) and fast for large values of $C_\mathrm{s}$ (small $\nicefrac{A}{C_\mathrm{s}}$). A special case is~\mbox{$A = C_\mathrm{s}$},~i.e.,~$\nacs = 1$, where all molecules are dissolved and released instantaneously when the carrier is placed into the medium. This is equivalent to the instantaneous release of molecules from a transparent sphere, which has been discussed in \cite{noel:16}.	

Based on assumptions A1) -- A3), the release process in Fig.~\ref{fig:1} can be described as a moving boundary problem~\cite{Higuchi:1963}. In particular, for a spherical homogeneous non-erodible matrix system with initial loading $A$ and solubility $C_\mathrm{s}$, the concentration $C(x,t)$ in the matrix can be described by the following partial differential equation~\cite[Eq.~(1)]{Lee_80}
\begin{align}
	\frac{\partial C}{\partial t} = x^{-2}\frac{\partial}{\partial x}\left(x^2 \Dm \frac{\partial C}{\partial x}\right),
	\label{eq:1}
\end{align}
where $\Dm$ is the diffusion coefficient of the molecules inside the matrix, which depends on the properties and structure of the polymer matrix.
As shown in Fig.~\ref{fig:1}, $x$ is the radial coordinate, the center of the sphere is at $x = 0$, and the surface at $x = a$.
% In \eqref{eq:1}, the actually 3D problem is reduced to a 1D problem referring to assumption A1.
Assuming equilibrium between the moving diffusion front and the environment, the boundary conditions are given by~\cite{Lee_80}
\vspace*{-1ex}
\begin{align}
	&C\big\vert_{x = a} = 0,\qquad
	&C\big\vert_{x = R(t)} = C_\mathrm{s}, 
%	\label{eq:2}\\[1ex]
	&&\Dm\frac{\partial C}{\partial x}\big\vert_{x = R(t)} &= (A - C_\mathrm{s}) \frac{\partial R}{\partial t}, \label{eq:3}
\end{align}
where $R(t)$ is the time-dependent position of the diffusion front, with initial position 	$R(0) = a$. 
%Moreover, the first condition in~\eqref{eq:3} describes the perfect sink condition.

\subsection{Drug Release from Diblock Copolymer Micelles}
\label{subsec:release_micelle}
\vspace*{-1ex}

% 
%In the following we introduce an example for the release of different drug molecules from a realistic drug carrier. 

In most studies of practical drug carriers, heuristic models are employed (cf., \cite{Macha_20, Siepmann2008, Arifin2006}), where measurement data are fitted to a mathematical model with several degrees of freedom. 
The reason why more elaborate models, e.g., based on the moving boundary problem \eqref{eq:1}, \eqref{eq:3}, are rarely applied is that it is cumbersome to estimate values for the parameters, e.g., $\nacs$, $\Dm$. 
Nevertheless, the application of more complex models is desirable, as they provide a direct relation between the amount of released drug molecules and the physical parameters of the drug and the carrier and, thus, give important insight into the release process. 
In \cite{Sutton2007}, the parameters for a model similar to the one discussed in Section~\ref{sec:drug}, based on the moving boundary problem \eqref{eq:1}, \eqref{eq:3}, have been estimated based on measurements of the release of different drug molecules from diblock copolymer micelles, acting as practical homogeneous matrix system. 

%\subsection{Drug Carrier and Drug Molecules}
\begin{table*}[t]
\caption{\small Parameter values for DOX and $\beta$-lap loaded in a micelle core \cite{Sutton2007}.}
\vspace*{-2ex}
\label{tab:acs}
\centering
\begin{tabular}{p{2.6cm}p{2cm}p{1cm}p{3cm}}
\hline\noalign{}
Drug molecule & $A$ in $\si{\milli\gram\per\milli\liter}$& $\nacs$ &$\Dm$ in $\si{\square\meter\per\second}$\\
\noalign{\smallskip}\hline\noalign{\smallskip}
Doxorubicin (DOX)  & &&\smallskip\\
pH~$5.0$  & $47.5$& $63.7$ & $1.82\cdot 10^{-22}$	\smallskip\\
pH~$7.4$ & $47.5$& $757.5$ & $1.82\cdot 10^{-22}$\\
\noalign{\smallskip}\hline\noalign{\smallskip}
$\beta$-lapachone ($\beta$-lap) & $14.1$ & $370.4$ & $2.42\cdot 10^{-21}$ 	\smallskip\\
\noalign{\smallskip}\hline\noalign{\smallskip}
\end{tabular}
\vspace*{-6ex}
\end{table*}

In the following, 
%we consider the practical drug carriers investigated in \cite{Sutton2007}.
%In particular,  
%we use the estimated parameter values in Section~\ref{sec:eval} to analyze the CR for practical drug carriers and drug molecules.
%
we consider diblock copolymer micelles, which are efficient drug carriers that have reached clinical trials in several cases \cite{Sutton2007}. 
Furthermore, micelles exhibit several properties beneficial for efficient DD, i.e., increased drug solubility, passive tumor targeting, and availability of several functionalizations \cite[Fig.~34-d]{Chude_Okonkwo_19}, \cite{Sutton2007, Ahmad2014}. 
A schematic illustration of a loaded diblock copolymer micelle is shown in Fig.~\ref{fig:micelle}.
The core of radius $a\approx 4.5\,\si{\nano\meter}$ is made of poly(DL-lactide) (PLA), in which the drug molecules are homogeneously distributed \cite{Sutton2007}. 
The core is surrounded by a polymer corona block made of poly(ethylene glycol) (PEG). 
The influence of the corona on the drug release kinetics is neglected for the derivation of the CR in this paper, i.e., we assume once a drug molecule is released from the core matrix of the diblock copolymer micelle, it is not further influenced by the corona and can diffuse freely in the channel. 

As drug molecules, we consider $\beta$-lapachone ($\beta$-lap) and doxorubicin (DOX), where DOX is one of the most common chemotherapeutic drugs \cite{Sutton2007,Weinberg2007}. 
The relevant model parameters for both drug molecules, estimated in \cite{Sutton2007}, are summarized in Table~\ref{tab:acs}. 
As the solubility $C_\mathrm{s}$ of DOX, in contrast to that of $\beta$-lap, exhibits a dependency on the pH value of the surrounding medium, different $\nacs$-values are considered to characterize the release of DOX at different pH levels. 
In particular, we consider a pH value of $5$ which is characteristic for tumor cells, while in blood vessels a pH of $7.4$ is typical \cite{D0BM01729A}.
From Table~\ref{tab:acs} we observe that the diffusion coefficients $\Dm$ are rather small as the drug molecules diffuse very slow inside the micelle core\footnote{Since these values have been estimated from real measurement data in \cite{Sutton2007}, they may comprise the influence of several physical and chemical effects that slow down the diffusion of drug molecules inside the micelle core.}.  
We note that there are several time-dependent effects occurring during the release process, e.g., DOX molecules interacting with the polymer, which further influence the release process. 
As previously mentioned, we neglect long-term interactions between the polymer and the \mbox{drug molecules in this paper}.

%% file: drug_release.tex
%\vspace*{-0.5ex}
In this section, based on the moving boundary problem in \eqref{eq:1}, \eqref{eq:3}, we present different models for the release of molecules from spherical homogeneous matrix systems, which serve as realistic TX models for the gradual release of molecules. In particular, we present two promising solutions from the literature on controlled drug release~\cite{Lee_80,Frenning:2004}, which are utilized as TX models in Section~\ref{sec:channel}. 
Moreover, we propose a simplified model for the molecule release from diblock copolymer micelles. 
Furthermore, for the validation of the theoretical results in Section~\ref{sec:eval}, we propose a PBS model for the gradual molecule release from matrix systems.
%
%
%investigate the molecule release from homogeneous matrix systems. 
%Depending on the structure (homogeneous or porous) of the matrix and the initial drug loading, several approximate analytical \cite{Lee_80, Higuchi:1963} and numerical \cite{Koizumi:1995} models have been derived, see \cite{Lee_11} for an overview.  
%\textcolor{red}{Revise. Hinweis auf ''not investigated in MC so far'', and focus on the TX! }

\subsection{Gradual Release Process}
\label{sec:drug_grad}
An analytical solution of the moving boundary problem~\mbox{\eqref{eq:1}, \eqref{eq:3}} has not been reported yet. 
However, a numerical solution has been presented in \cite{Koizumi:1995}, where the finite difference method~(FDM) is used to solve the moving boundary problem, which provides very accurate results.  
Therefore, we use the solution from \cite{Koizumi:1995} as ground truth for the numerical evaluation in Section~\ref{sec:eval}. 
However, the FDM entails high computational complexity and, thus, different approximate analytical solutions have been presented in \cite{Lee_80,Higuchi:1963,Koizumi:1995, Frenning:2004}. 
All these solutions have in common that their accuracy is rather poor for small~$\nacs$ values  (e.g., $\nacs < 10$), corresponding to a fast release process. For larger $\nacs$ values, which are typically of interest in practice (cf.~Section~\ref{subsec:release_micelle}), the accuracy of the approximate solutions improves~\cite{Lee_11}.
 
A promising solution is presented in \cite{Lee_80},
% as it is valid for the widest range of $\nacs$ ratios.
and it is obtained by normalization of \eqref{eq:1}, \eqref{eq:3}, and the application of a \textit{double-integration heat balance integral}\footnote{The individual steps of the solution are not shown here for brevity, but are given in \cite{Lee_80}.} approach \cite{Langford:1971}. 
This results in an analytical expression for the amount of molecules $M$ that are absorbed at the surface of the matrix~(perfect sink), which is equal to the amount of molecules released into the surrounding medium at $x = a$. 
The fraction of released molecules can be expressed as a function of the normalized diffusion front position~$\delta = 1 - \nicefrac{R(t)}{a}$ and is given by~\cite[Eq.~(28)]{Lee_80}
%\vspace*{-1ex}
\begin{align}
	\frac{M(\delta)}{M_\infty}&\bigg\vert_{\acs \geq 1} \!\!= \left[1 -(1-\delta)^3 \right]\left(1 - \csa \right)  + 
%	\nonumber\\
%	&
	3\delta\csa\left[\left( a_1 + \frac{a_2}{2} + \frac{a_3}{3}\right) - \left(\frac{a_1}{2} + \frac{a_2}{3} +\frac{a_3}{4}\right)\delta\right]\!, \label{eq:9} 
\end{align}
where $M_\infty$ is the total amount of available molecules and coefficients $a_1, a_2$, and $a_3$ are given by \cite[Eqs.~(19), (19a)]{Lee_80} %$M_\infty = A \frac{4}{3}\pi a^3
\vspace*{-1ex}
\begin{align}
	&a_1 = 1, &a_2 = -a_3 -1,
%	\label{eq:11}\\
	&&a_3 = \lambda - \sqrt{\lambda^2 - 1}, 
	&&\lambda = 1-\left( 1 - \acs\right)(1-\delta) \label{eq:12}.
\end{align}
We note that the result in \eqref{eq:9} assumes that dissolved molecules are reflected at the undissolved core of the matrix. Moreover, when the diffusion front reaches~$R(t) = 0$ all molecules are dissolved and \eqref{eq:9} should yield one for~$\delta = 1$, but it gives $\nicefrac{M(1)}{M_\infty} = 1 - \nicefrac{C_\mathrm{s}}{(4A)}$. This confirms the previously mentioned approximate character of \eqref{eq:9} \cite{Lee_80}, where the accuracy increases with increasing~$\nacs$. 
Complementary to~\eqref{eq:9}, the normalized position of the diffusion front $\delta$ can be expressed as a  function of time as follows \cite[Eq.~(25)]{Lee_80}
\begin{align}
	\frac{\Dm t}{a^2} = \frac{1}{12}\left[6\frac{A}{C_\mathrm{s}} - 4 -a_3\right]\delta^2 - \frac{1}{3}\left(\acs - 1\right)\delta^3. \label{eq:10}
\end{align}
%The coefficients $a_1, a_2$, and $a_3$ in \eqref{eq:9}, \eqref{eq:10} follow 
%%from the a polynomial approximation of normalized concentration $\theta$ and can be derived 
%as \cite[Eqs.~(19), (19a)]{Lee_80}
%\begin{align}
%	a_1 &= 1, &a_2 &= -a_3 -1,\label{eq:11}\\
%	a_3 &= \lambda - \sqrt{\lambda^2 - 1}, 
%	&\lambda &= 1-\left( 1 - \acs\right)(1-\delta) \label{eq:12}.
%\end{align}
Setting $\delta = 1$ in \eqref{eq:10} yields the time duration $t_\mathrm{rel}$ needed for all molecules to be released from the matrix 
\begin{align}
t_\mathrm{rel} = \frac{a^2}{\Dm}\left(\frac{1}{6}\acs - \frac{1}{12}\right). \label{eq:release_time}
\end{align}
%For practical application it would be desirable to have a single function that describes the drug release over time. 
To describe the amount of molecules released over time, \eqref{eq:9} and \eqref{eq:10} have to be evaluated simultaneously. 
Exploiting that the diffusion front $\delta$ ranges from $0$ to $1$, i.e., $R(0) = a$ and $R(t_\mathrm{rel}) = 0$, allows the calculation of the number of released molecules $M$ at time $t$ based on \eqref{eq:9} and \eqref{eq:10}. 
We note that \eqref{eq:9} and \eqref{eq:10} are only valid for $t\in[0,t_\mathrm{rel}]$, because $t > t_\mathrm{rel}$ would correspond to a negative position of the diffusion front $R(t)$, which is physically not possible.

For the design of MC systems using a matrix system as practical TX model, it is desirable to describe the amount of released molecules $\nicefrac{M}{M_\infty}$ in \eqref{eq:9} as an explicit function of time. 
In \cite{Frenning:2004}, an approximate function for the  release over time has been proposed for large values of $\nacs$. 
In particular, assuming~$\nacs \gg 1$ allows to simplify~\eqref{eq:10} such that it can be solved for the diffusion front position $R(t)$ in terms of a cubic root~\cite[Sec.~2.3]{Frenning:2004}
\begin{align}
	\frac{R}{a} = &\frac{1}{2}\left(1 - \frac{1}{3}\csa\right)+\! \left(1 \!+\! \frac{1}{3}\csa\right) \!\cos\!\left(\frac{\arccos\left(12\frac{C_\mathrm{s}\Dm}{A a^2}t - 1\right) + 4\pi}{3}\right)\!, \label{eq:frenning:delta}
\end{align} 
$t\in [0, t_\mathrm{rel}]$. Applying the same assumption, the normalized number of released molecules \eqref{eq:9} can be simplified as follows \cite{Frenning:2004}
\begin{align}
	\frac{M(t)}{M_\infty}\bigg\vert_{\acs \gg 1} \!\!\!\!\!\!= 1 \!-\! \left(\frac{R}{a}\right)^3 \!\!\!+ \frac{1}{2}\frac{C_\mathrm{s}}{A}\!\left[2\left(\frac{R}{a}\right)^3 \!\!\!-\! \left(\frac{R}{a}\right)^2 \!\!\!-\! \frac{R}{a} \right]\!\!, \label{eq:frenning:M}
\end{align}
$t\in [0, t_\mathrm{rel}]$. An explicit closed-form expression for the normalized number of released molecules as a function of time $t$ can be obtained by inserting \eqref{eq:frenning:delta} into \eqref{eq:frenning:M}. 
The validity of the simplified model \eqref{eq:frenning:delta}, \eqref{eq:frenning:M} is discussed in Section~\ref{sec:eval}.

\subsection{Instantaneous Release Process}
\label{sec:drug_ins}
\vspace*{-1ex}
Next, we consider the special case $\nacs = 1$, where the molecule solubility $C_\mathrm{s}$ is equal to the initial loading $A$. 
This means the diffusion front immediately reaches the center of the matrix. 
This is equivalent to the case, where molecules are distributed uniformly over the entire volume of a transparent sphere, and are then instantaneously released.
As the release process is no longer a moving boundary problem, simpler mathematical descriptions can be used instead.
%\footnote{We note that, for the case of $\nacs = 1$, the ratio is no longer relevant for the release process, and therefore it will not occur in the results.}.

The CR for such a transparent spherical TX was numerically derived in~\cite{noel:16}, based on the well-known point TX model. 
However, no closed-form expression for the number of molecules released from the TX has been provided. 

The number of molecules released from a sphere, uniformly filled with molecules, into a bounded release medium is given in~\cite[Eq.~(6.30)]{crank:1975}. This solution can be extended to an unbounded release medium, which corresponds to the normalized number of released molecules of a matrix for the special case $\mbox{$\nacs = 1$}$, as follows~\cite{Frenning:2004} 
\begin{align}
	\frac{M(t)}{M_\infty}\bigg\vert_{\acs = 1} = 1 - \frac{6}{\pi^2}\sum\limits_{n=1}^\infty \frac{1}{n^2}\mathrm{exp}\left(\gamma_n t\right),
	 \label{eq:crankVinf1}
\end{align}
where $\gamma_n = -\Dm n^2\frac{\pi^2}{a^2}$. In the context of drug release from homogeneous matrix systems, \eqref{eq:crankVinf1} has also been reported in \cite{Arifin2006}.
%\footnote{
We note that \eqref{eq:crankVinf1} can also be applied for $\nacs < 1$ because the molecule release is still instantaneous if the solubility $C_\mathrm{s}$ is larger than the initial loading $A$ \cite{Arifin2006}.

\subsection{Release Process Model for Diblock Copolymer Micelles}
\label{subsec:model_micelle}
\vspace*{-1ex}

From the practical parameter values in Table~\ref{tab:acs} we observe that the $\nacs$-values for DOX and $\beta$-lap are very large, while the diffusion coefficients $\Dm$ inside the matrix are very small. 
Inserting these values into \eqref{eq:release_time}, leads to very large durations, in the order of hours and days, until all drug molecules are released.
%\footnote{These very large release durations originate from the fact that in practice only a small percentage of all loaded drugs is released from the matrix carrier \cite{Sutton2007}.}.  
%
The large $\nacs$-values of practical drug molecules allow to introduce further simplifications for the molecule release model in \eqref{eq:frenning:delta} and \eqref{eq:frenning:M}. 
In particular, both \eqref{eq:frenning:delta} and \eqref{eq:frenning:M} contain terms with $\nicefrac{C_\mathrm{s}}{A}$ which is close to zero for practical $\nacs$-values. 
Hence, we can simplify \eqref{eq:frenning:delta} as follows 
\begin{align}
	\frac{R}{a} = &\frac{1}{2}+\sin\!\left(\frac{\arcsin\left(1 - 12\frac{C_\mathrm{s}\Dm}{A a^2}t\right)}{3}\right)\!, \label{eq:frenning:delta_v2}
\end{align} 
where the $\cos$-functions in \eqref{eq:frenning:delta} have been rewritten in terms of $\sin$-functions. 
%together with the $4\pi$ phase modulation occurring in \eqref{eq:frenning:delta}. 
Similarly, \eqref{eq:frenning:M} can be further simplified for large $\nacs$-values as follows 
\begin{align}
	\frac{M(t)}{M_\infty}\bigg\vert_{\acs \gg 1} \!\!\!\!\!\!= 1 \!-\! \left(\frac{R}{a}\right)^3 
%	= 1 - \left[\frac{1}{2}+\sin\!\left(\frac{\arcsin\left(1 - 12\frac{C_\mathrm{s}D}{A a^2}t\right)}{3}\right) \right]^3
	. \label{eq:frenning:M_v2}
\end{align}
The molecule release models in \eqref{eq:frenning:delta} and \eqref{eq:frenning:M} have been derived from the presented general model in \eqref{eq:9}, \eqref{eq:10} under the assumption that $\nacs \gg 1$. 
Furthermore, based on the large $\nacs$-values for practical drug molecules in Table~\ref{tab:acs}, \eqref{eq:frenning:delta_v2} and \eqref{eq:frenning:M_v2} have been derived. 
From our investigations, we found that these models are accurate for realistic values of $\nacs \geq 100$. The validity of these simplifications is shown in Section~\ref{sec:eval}.

\subsection{Particle-Based Simulation}
\label{sec:drug_pbs}
%\textcolor{red}{Extend this section}

% ----- HACK-ME -----
\begin{algorithm}[t!]
\SetAlgoLined
\DontPrintSemicolon
% Determine diffusion front $R(t_k)$ using FDM \cite{Koizumi:1995} \\
Define sphere of radius $a$ and center point $\mathbf{r}_\text{s}$\;\vspace*{-1ex}
Distribute $M_\infty$ molecules uniformly in the sphere $\to \bm{r}_m(t_1)$\;\vspace*{-1ex}
Pre-calculate diffusion front positions $\to R(t_k)$\;\vspace*{-1ex}
Initialize number of released molecules $M(t_1) = 0$ 
\;\vspace*{-1ex}
 \For{$k \leftarrow$ 1 \KwTo $K$ }{ \vspace*{-1ex}
  \For{$m \leftarrow$ 1 \KwTo $M_\infty$}{\vspace*{-1ex}
   \If{molecule is not marked as released}{\vspace*{-1ex}
   $d_m(t_k) = ||\mathbf{r}_m(t_k) - \mathbf{r}_\text{s}||_2$\;\vspace*{-1ex}	
     \If{$d_m(t_k) \geq R(t_k)$}{\vspace*{-1ex}
     $\mathbf{r}_m(t_{k+1}) =\mathbf{r}_m(t_k) + \mathcal{N}(\mathbf{0},2\Dm\Delta t \mathbf{I})$ \;\vspace*{-1ex}
     $d_m(t_{k+1}) = ||\mathbf{r}_m(t_{k+1}) - \mathbf{r}_\text{s}||_2$\;\vspace*{-1ex}
       \If{$d_m(t_{k+1}) < R(t_k)$}{\vspace*{-1ex}
       $\mathbf{r}_m(t_{k+1}) = \mathbf{r}_m(t_k)$\tcp*{\footnotesize reflection at undissolved matrix core}\vspace*{-1ex}
       }\vspace*{-1ex}
     }\vspace*{-1ex}
     } \vspace*{-1ex}
     \If{$d_m(t_k) \geq a$}{\vspace*{-1ex}
     $M(t_k) = M(t_k) + 1$ \;\vspace*{-1ex}
     Mark molecule as released\vspace*{-1ex}
     }\vspace*{-1ex}
   }\vspace*{-1ex}
   $M(t_{k+1}) = M(t_k)$\vspace*{-1ex}
 }\vspace*{-1ex}
 \caption{PBS to determine the number of molecules released from spherical matrix (see Section~\ref{sec:drug_pbs})}
 \label{alg:pbs}
\end{algorithm}
% ----- HACK-ME -----

To validate the expressions for the number of molecules released from matrix systems presented in \eqref{eq:9}, \eqref{eq:frenning:M}, and~\eqref{eq:crankVinf1}, we have developed a PBS model. 
The simulator was implemented in the programming language Python and time-consuming parts were realized using Cython. 
The individual simulation steps are summarized in Algorithm~\ref{alg:pbs}. 
The simulation takes into account reflections at the undissolved matrix core, by assuming that a reflected molecule bounces back to its previous position~\cite{Deng_16} (see line $13$ in Algorithm~\ref{alg:pbs}). 
This is in line with the assumptions made for the derivation of the theoretical results presented above. 
After the simulation, the results from multiple simulation runs are accumulated and averaged.
%The simulation algorithm is executed in parallel and after the simulation, the results from different simulation runs are accumulated and averaged.

%To keep the presentation of Algorithm~\ref{alg:pbs} concise, we used the following definitions: 
The time step is denoted by~$\Delta t$ and the discrete time instances are given by $t_k=k\Delta t$ with a maximum number of time steps $K$. 
The position of the $m$th molecule at time $t_k$ is given by~$\mathbf{r}_m(t_k) = \left[x_m(t_k)\, y_m(t_k)\, z_m(t_k) \right]\tran$ and its Euclidean distance to the origin of the sphere $\mathbf{r}_\text{s}$ is calculated as $d_m(t_k) = ||\mathbf{r}_m(t_k) - \mathbf{r}_\text{s}||_2$.
A molecule starts to diffuse when it is dissolved by the diffusion front, i.e., when $d_m(t_k) \geq R(t_k)$. 
The diffusion front position $R$ at times $t_k$ can be pre-calculated, e.g., by FDM \cite{Koizumi:1995}.
The molecule movement is simulated as a random walk, where the molecule position is updated as $\mathcal{N}(\mathbf{0},2\Dm\Delta t \mathbf{I})$. 
Here, $\mathcal{N}(\pmb{\mu}, \pmb{\Sigma})$ denotes a multivariate Gaussian distribution with mean vector $\pmb{\mu}$ and covariance matrix $\pmb{\Sigma}$, and $\mathbf{0}$ and $\mathbf{I}$ denote the all-zero vector and the identity matrix, respectively. 

%The simulation algorithm initially distributes the molecules uniformly over the entire volume (i.e., all molecules are undissolved). Then, the molecules are gradually released, with the individual simulation steps summarized in Algorithm~\ref{alg:pbs}. In particular, the procedure for each molecule at each time step $\Delta t$ is shown. The diffusion front position $R(t)$ is determined in advance at times~$t_k=k\Delta t$ using the FDM~\cite{Koizumi:1995}. Similar to the presented theoretical expression, the PBS considers reflections at the undissolved matrix core by undoing the last movement~\cite{Deng_16}. 

%% file: channel.tex
\begin{figure}[!t]	
	\centering
	\includegraphics[width=0.5\linewidth]{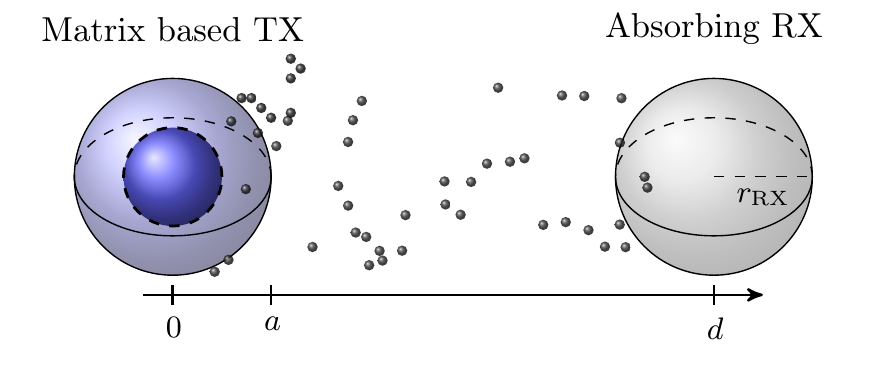}
	\vspace*{-2ex}
	\caption{\small System model for the molecule release from a spherical matrix in a 3D unbounded environment with an absorbing RX.}
	\label{fig:system}
	\vspace*{-3ex}
\end{figure}

In Section~\ref{sec:drug}, the gradual and instantaneous release of molecules from a spherical matrix have been analyzed, assuming a perfect sink condition at the matrix boundary~(see~\eqref{eq:3}). 
In the following, we extend this study and derive the CR for a spherical matrix TX and an absorbing spherical RX in free space, see Fig.~\ref{fig:system}.

The models presented in Sections~\ref{sec:drug_grad} -- C
%\ref{subsec:model_micelle}
 characterize the normalized number of molecules released at the matrix surface at $x = a$. Therefore, the~CR can be derived by combining the molecule release of the matrix system and the instantaneous release of molecules from the surface of a transparent spherical TX. 
For the following analysis, we assume that once molecules are released into the surrounding medium the matrix does not impede the diffusion of the molecules, i.e., reflections at the TX are neglected\footnote{Although assuming a reflective TX would be more realistic, it has been shown in \cite{noel:16} that a transparent TX surface is an appropriate approximation.}. 

\subsection{Instantaneous Release from a Surface Transmitter}

Assuming instantaneous release of the molecules from the surface of a sphere at $t = 0$, the hitting probability of the molecules at an absorbing RX of radius $r_\mathrm{RX}$ is given by \cite[Eq.~(9)]{Huang2020}
\begin{align}
p_\mathrm{s}(t)&=
\frac{2\rho a \rrx}{d}
\sqrt{\frac{\pi \Dc}{t}}
\bigg[
\mathrm{exp}{\left(\!-\frac{\beta_1}{t} \right)} \!-\! 
\mathrm{exp}{\left(\!-\frac{\beta_2}{t} \right)} 
\bigg]\!,
\label{TS_TX_Abs_RX,Rate}
%\\
%&= f \sqrt{\frac{1}{t}} 
%\bigg[\mathrm{exp}\bigg(-\frac{g}{t}\bigg) - 
%\mathrm{exp}\bigg(-\frac{h}{t}\bigg) \bigg]
\end{align}
where $\Dc$ is the diffusion coefficient in the release medium, and $\beta_1 =\frac{ (a+\rrx)(a+\rrx-2d) + d^2 }{4\Dc}$ and $\beta_2 = \frac{ (a-\rrx)(a-\rrx+2d) + d^2 }{4\Dc}$. Here, the distance between the centers of the TX and RX is denoted by $d$ (see Fig.~\ref{fig:system}), and $\rho = \left(4 \pi a ^2\right)^{-1}$. 
%We note that \eqref{TS_TX_Abs_RX,Rate} slightly differs from \cite[Eq.~(9)]{Huang2020}, because we do not consider the effect of membrane fusion, i.e., $k_\mathrm{f} = 0$ in \cite[Eq.~(9)]{Huang2020}. 
%A proof for \eqref{TS_TX_Abs_RX,Rate} is given in \cite[Appendix~B]{Huang2020}. 
From the hitting probability \eqref{TS_TX_Abs_RX,Rate}, the amount of molecules absorbed by the RX up to time $t$ can be obtained by integration 
\begin{align}
	N_\mathrm{s}(t) &= \int_0^t p_\mathrm{s}(\xi)\dint{\xi} = \frac{4\rho a \rrx}{d}\sqrt{\pi \Dc}\left\lbrace \sqrt{t}\left[\expI{-\frac{\beta_1}{t}} - \expI{-\frac{\beta_2}{t}}\right] \right.\nonumber\\
	&\left.-\sqrt{\pi}\left[\sqrt{\beta_2}\erf{\sqrt{\frac{\beta_2}{t}}}  - \sqrt{\beta_1}\erf{\sqrt{\frac{\beta_1}{t}}}\right] + \sqrt{\pi} \left(\sqrt{\beta_2} - \sqrt{\beta_1}\right)\right\rbrace .
	\label{eq:TrSurfTX_N}
\end{align}

\subsection{Channel Response}

The overall CR for the release of molecules from a matrix system and an absorbing RX in free space can be obtained by the convolution of the amount of molecules released over time, $M(t)$, and the hitting probability, $p_\mathrm{s}(t)$, from \eqref{TS_TX_Abs_RX,Rate} as follows
\begin{align}
	N(t) = p_\mathrm{s}(t) \ast M(t) = \int_0^t p_\mathrm{s}(t - \xi)M(\xi)\dint{\xi} = N_\mathrm{s}(t)\ast \frac{\mathrm{d}}{\mathrm{d}t}M(t), 
	\label{eq:cr:conv1}
\end{align} 
%Utilizing the amount of absorbed molecules $N_\mathrm{s}(t)$ in \eqref{eq:TrSurfTX_N}, an alternative expression for \eqref{eq:cr:conv1} can be obtained by the convolution with the drug release rate $m(t)$ as follows
%\begin{align}
%	N(t) = m(t)\ast N_\mathrm{s}(t) = \int_0^t m(t-\xi)N_\mathrm{s}(\xi)\dint{\xi}, \label{eq:cr:conv2}
%\end{align}
%where $m(t) = \frac{\partial}{\partial t} M(t)$ is the molecule release rate of the matrix TX. 
%Both expressions \eqref{eq:cr:conv1} and \eqref{eq:cr:conv2} are equivalent, as for the convolution holds $\frac{\partial}{\partial t}x(t)\ast y(t) = \frac{\partial}{\partial t}y(t)\ast x(t)$. 
where $f(t)\ast g(t) = \int_0^t f(t-\xi)g(\xi)\dint{\xi}$ denotes a convolution with respect to time. In the following, we simplify \eqref{eq:cr:conv1} for the cases $\acs \gg 1$ and $\acs = 1$. 

\subsubsection{Gradual Release}
\label{subsec:gradual}

To obtain the CR for the gradual release of molecules from a matrix system, i.e., $\nacs \gg 1$, we have to extend the approximate solutions in \eqref{eq:frenning:M} and \eqref{eq:frenning:M_v2} to the interval $t\in [0,\infty)$. Exploiting that the normalized number of released molecules $\nicefrac{M}{M_\infty}$ should remain~$1$ when~$t \geq t_\mathrm{rel}$, we express the amount of molecules released from the matrix surface as follows
\begin{align}
	\bar{M}(t)\big\vert_{\acs\gg1} \!=\! M(t)\big\vert_{\acs\gg1}\!\left(\epsilon(t) - \epsilon(t - t_\mathrm{rel})\right) + M_\infty \epsilon(t - t_\mathrm{rel}),
	\label{eq:mcon}
\end{align}
where $\epsilon(t)$ denotes the unit step function. 
Inserting \eqref{eq:mcon} into \eqref{eq:cr:conv1} leads to the CR for a gradual release from a matrix TX as follows 
%To obtain the number of molecules absorbed at the RX, \eqref{eq:mcon} has to be convolved with the hitting probability \eqref{TS_TX_Abs_RX,Rate}, i.e.,
\vspace*{-0.5ex}
\begin{align}
	N(t)\big\vert_{\acs\gg1} = \int_0^t p_\mathrm{s}(t - \xi)\bar{M}(\xi)\big\vert_{\acs\gg1} \,\mathrm{d}\xi.
	\label{eq:response_con}
\end{align}
In particular, \eqref{eq:response_con} specifies the number of molecules absorbed at the RX for a gradual release of molecules at the TX. 
Due to the complex structure of $M(t)$ (see \eqref{eq:frenning:M} or \eqref{eq:frenning:M_v2}) a closed-form solution for~\eqref{eq:response_con} could not be found. 
However, to study the influence of a gradual release of molecules from the matrix system on the CR, only the one-dimensional integral in \eqref{eq:response_con} has to be evaluated numerically (see Section~\ref{sec:eval}).

\subsubsection{Instantaneous Release Process}
\label{subsec:inst}

To obtain the CR for an instantaneous release from the matrix system ($\nacs = 1$), we insert \eqref{eq:crankVinf1} into \eqref{eq:cr:conv1}, yielding
\vspace*{-0.5ex}
\begin{align}
	N(t)\big\vert_{\acs=1} = \int_0^t p_\mathrm{s}(t - \xi)M(\xi)\big\vert_{\acs=1} \,\mathrm{d}\xi. 
	\label{eq:response_ins1}
\end{align}
The evaluation of the integral leads to an analytical expression for $N(t)$ that is given in \eqref{eq:response_ins2}, shown on top of this page, where $\mathrm{erf}(x)$ and $\mathrm{erfc}(x)$ denote the error function and the complementary error function, respectively.
We note that \eqref{eq:response_ins2} specifies the number of absorbed molecules in response to an instantaneous release from a spherical matrix with $\nacs = 1$. 
This is equivalent to the instantaneous release from a transparent sphere, which was considered in \cite{noel:16}, but where no analytical expression for $N(t)$ was provided. For the numerical evaluation of CR \eqref{eq:response_ins2} in Section~\ref{sec:eval} we used $n = 1, \dots, 300$ terms to approximate the infinite sum. 
\begin{figure*}[!t]
% ensure that we have normalsize text
\normalsize
% Store the current equation number. \setcounter{MYtempeqncnt}{\value{equation}}
% Set the equation number to one less than the one
% desired for the first equation here.
% The value here will have to changed if equations
% are added or removed prior to the place these
% equations are referenced in the main text.
\setcounter{equation}{17}
\small
\begin{align}
N(t)\bigg\vert_{\acs=1} \!\! \!\!\!\!\!&=\!
\frac{M_\infty  r_{\mathrm{RX}}}{\pi a  d} \sqrt{\pi \Dc} \bigg[
\sqrt{t} \bigg( \mathrm{exp}\bigg(\!\!\!-\!\frac{\beta_1}{t} \bigg) \!-\!
\mathrm{exp}\bigg(\!\!\!-\!\frac{\beta_2}{t} \bigg) \bigg)
\!+\!
\sqrt{\pi}
\bigg(\sqrt{\beta_1}\mathrm{erf}\bigg(\!\!\sqrt{\frac{\beta_1}{t}} \bigg) \!-\!
\sqrt{\beta_2}\mathrm{erf}\bigg(\!\!\sqrt{\frac{\beta_2}{t}} \bigg)
\!+\! \sqrt{\beta_2} \!-\! \sqrt{\beta_1}
\bigg)
\!\bigg]
\notag
\\[-0.5ex]
&+
\sum _{n=1}^\infty
\frac{3{M_\infty} r_{\mathrm{RX}} \sqrt{\Dc}}
{2d a n^2 \pi^2 \sqrt{\gamma_n}} 
%\mathrm{exp}\left(\gamma_n t\right) 
\bigg\{ 
\mathrm{exp}\bigg(\!\gamma_n t-\!2 \sqrt{\gamma_n\beta_1} \bigg)
\bigg[
\mathrm{exp}\bigg(4 \sqrt{\gamma_n\beta_1} \bigg)
\cdot
\mathrm{erfc}\bigg(\sqrt{\gamma_n t} \!+\! \sqrt{\frac{\beta_1}{t}} \bigg) +
\mathrm{erfc}\bigg(\sqrt{\frac{\beta_1}{t}} \!-\! \sqrt{\gamma_n t}  \bigg) 
\bigg]
\notag
\\[-0.5ex]
&-
\mathrm{exp}\bigg(\!\gamma_n t- \!2 \sqrt{\gamma_n\beta_2} \bigg)
\bigg[
\mathrm{exp}\bigg(4 \sqrt{\gamma_n\beta_2} \bigg)
\cdot
\mathrm{erfc}\bigg(\sqrt{\gamma_n t} \!+\! \sqrt{\frac{\beta_2}{t}} \bigg) +
\mathrm{erfc}\bigg( \sqrt{\frac{\beta_2}{t}} \!-\! \sqrt{\gamma_n t}  \bigg) 
\bigg]
\bigg\}
\label{eq:response_ins2}
\end{align}
% Restore the current equation number. 
\setcounter{equation}{18}
% The IEEE uses as a separator
\vspace*{-0.5ex}
\hrulefill
% The spacer can be tweaked to stop underfull vboxes. \vspace*{4pt}
\vspace*{-5ex}
\end{figure*}

%% file: regimes.tex
The CR for a matrix TX and an absorbing RX depends on both the dynamics of the molecule release process $M(t)$ from the matrix TX and the properties of the diffusive channel, characterized by $p_\mathrm{s}(t)$ or $N_\mathrm{s}(t)$, between TX and RX. 
In the following, we first derive the duration $\tabs$ until a released molecule is absorbed by the RX, which serves as a measure for the dynamics of the channel.
Then, together with the duration of the release process $\trel$, we define a criterion for determining whether the release process or the channel dynamics are more important for the characteristics of the CR. 
Based on this criterion, we define two limiting regimes, i.e., the \textit{channel dominated regime} and the \textit{release dominated regime}, where the channel dynamic or the release process are more important, respectively.
Finally, we propose two closed-form expressions that approximate the CR in the limiting regimes.

\subsection{Duration of the Molecule Absorption Process}
In principle, it takes infinitely long until an amount of $\frac{\rrx}{d}$ molecules is absorbed by the RX,
%, i.e., $\tabs\to\infty$
which can be shown by taking the limit value of \eqref{eq:TrSurfTX_N}, i.e., $\lim_{t\to\infty}N_\mathrm{s}(t) = \frac{\rrx}{d}$.
However, the amount of molecules absorbed by the RX decreases significantly for large times as the hitting probability  \eqref{TS_TX_Abs_RX,Rate} tends to zero for large times, i.e., $\lim_{t\to\infty}p_\mathrm{s}(t) = 0$ (see Fig.~\ref{fig:absRate}).
Therefore, we define $\tabs$ as the non-infinite duration until a high percentage of molecules is absorbed by the RX, i.e., $\frac{N_\mathrm{s}(\tabs)}{N_\mathrm{s}(t\to\infty)} = \sigma$, where $N_\mathrm{s}(\tabs) \approx N_\mathrm{s}(t\to\infty)$ for values of $\sigma$ close to~$1$.  
For the considered transparent surface TX, $\tabs$ can be obtained from $N_\mathrm{s}$ in \eqref{eq:TrSurfTX_N} by solving $N_\mathrm{s}(\tabs) = \sigma\frac{\rrx}{d}$ for $\tabs$, which can only be done numerically.

Since we want to derive an analytical expression that reveals the dependencies of the duration $\tabs$ on characteristic channel parameters, e.g., $\Dc$ and $d$, we exploit that the amount of molecules absorbed by a transparent surface TX, $N_\mathrm{s}$, and that of a point TX, $N_\mathrm{p}$, become identical for $t \to \infty$, i.e., $N_\mathrm{s}(t\to\infty) = N_\mathrm{p}(t\to\infty) = \frac{\rrx}{d}$ (see Fig.~\ref{fig:3} and \cite[Figs.~1, 2]{noel:16}). 
Hence, we calculate $\tabs$ based on the amount of absorbed molecules for a point TX and an absorbing RX, which is given by \cite[Eq.~(6)]{noel:16}
\begin{align}
	N_\mathrm{p}(t) = \frac{\rrx}{d}\erfc{\frac{d - \rrx}{\sqrt{4\Dc t}}}. \label{eq:Point_N}
\end{align}
Then, the duration $\tabs$ until an amount of $\sigma\frac{\rrx}{d}$ molecules are absorbed by the RX can be obtained from \eqref{eq:Point_N} as follows 
\begin{align}
	&N_\mathrm{p}(\tabs)= \sigma\frac{\rrx}{d}= \frac{\rrx}{d}\erfc{\frac{d - \rrx}{\sqrt{4\Dc \tabs(\sigma)}}} &\leadsto &&\tabs(\sigma) = \frac{(d - \rrx)^2}{4\Dc}\frac{1}{\left(\erfcInv{\sigma} \right)^2},\label{eq:tabs:2}
\end{align}
where $\erfcInv{x}$ is the functional inverse of $\erfc{x}$. 
The duration $\tabs$ in \eqref{eq:tabs:2} can also be related to the peak time of the hitting probability of a point TX, $t_\mathrm{peak, p}$, as follows 
\begin{align}
	&\tabs(\sigma) = \frac{3}{2}t_\mathrm{peak, p}\frac{1}{\left(\erfcInv{\sigma} \right)^2}, &\mathrm{with}
	&&t_\mathrm{peak, p} = \frac{(d - \rrx)^2}{6\Dc}.\label{eq:tabs:3}
\end{align} 
The derivation of the peak time $t_\mathrm{peak, p}$ is discussed in \cite{noel:16}. 
We note that although the duration $\tabs$ was derived for a point TX \eqref{eq:Point_N}, it is also a good approximation for the considered transparent surface TX, 
%for large $\sigma$-values, 
as both become identical for large $t$, i.e., $N_\mathrm{p}(\tabs) \approx N_\mathrm{s}(\tabs)$ for $\sigma$ close to $1$.

\subsection{Relative Importance of Release and Channel Dynamics} 

To obtain a criterion that determines whether the release process or the channel dynamics are more important for the characteristics of the CR, we utilize the ratio between the duration of the molecule release process from the matrix TX $\trel$ in \eqref{eq:release_time}, and the duration until the molecules are absorbed by the RX $\tabs$ in \eqref{eq:tabs:3}, as follows 
\begin{align}
	\ratio = \frac{\trel}{\tabs} = 
	\frac{2a^2}{3(d - \rrx)^2}\frac{\Dc}{\Dm}\left(\acs - \frac{1}{2}\right)\left(\erfcInv{\sigma}\right)^2. \label{eq:ratio:2}
\end{align}
We note that \eqref{eq:ratio:2} is only valid for $\nacs > 1$, as for $\nacs \leq 1$ all molecules are released instantaneously, i.e., $\trel = 0$, yielding $\tau = 0$.
The ratio $\tau$ in \eqref{eq:ratio:2} can be exploited to determine whether the molecule release process or the channel dynamics are more important for the characteristics of the overall CR 
and allows us to define two limiting regimes for a DD process:
\begin{itemize}
	\item \textit{Channel dominated regime} ($\ratio \ll 1$): 
	The overall CR is determined by the channel dynamics, while the impact of the exact dynamics of the release process is negligible. 
%	According to \eqref{eq:ratio:2}, this regime is reached for, e.g., large distances $d$ between TX and RX, small channel diffusion $\Dc$, and small $\nacs$ values in the case of a matrix TX. 
	\item \textit{Release dominated regime} ($\ratio \gg 1$): 
	The overall CR is determined by the release process, while the impact of the exact channel dynamics is negligible. 
%	According to \eqref{eq:ratio:2}, this regime is reached for, e.g., small distances $d$ between TX and RX, and for a slow release of molecules from the TX due to a small $\Dm$ or large $\nacs$ for matrix type TXs.   
\end{itemize} 
For both limiting regimes, the derivation of the overall CR \eqref{eq:cr:conv1}, which has to be solved numerically in general (see Section~\ref{sec:channel}), can be simplified. 
These simplifications for the \textit{channel} and the \textit{release dominated regimes} and the resulting closed-form expressions for the CR are discussed in the following.

\subsection{Channel Dominated Regime}

First, we investigate the \textit{channel dominated regime}, i.e., $\trel \ll \tabs$ and $\ratio \ll 1$. 
Here, the dynamic of the channel is much slower than that of the release process and determines the overall CR while the molecule release process has less influence. 
Inspecting \eqref{eq:ratio:2}, this regime is reached  for small $\nacs$-values, corresponding to a very fast molecule release, for large distances $d$ between TX and RX, and for small diffusion coefficients $\Dc$, respectively.

In this regime, the exact dynamics of the release process do not contribute to the overall CR in \eqref{eq:cr:conv1}, and thus, the gradual release of molecules can be approximated by an instantaneous release. 
Therefore, the number of molecules absorbed by the RX can be obtained by the convolution of hitting probability $p_\mathrm{s}$ with a unit step function weighted with the number of molecules $M_\infty$ released from the matrix TX as follows 
\begin{align}
	N(t) = p_\mathrm{s}(t) \ast M(t) \approx M_\infty \epsilon(t) \ast p_\mathrm{s}(t) = M_\infty \int_0^{t}p_\mathrm{s}(\tau)\dint{\tau}
= M_\infty N_\mathrm{s}(t) = \tilde{N}(t)\big\vert_{\ratio \ll 1},
\label{eq:cr:beta-}
\end{align}
where $N_\mathrm{s}(t)$ is the amount of absorbed molecules at the RX in \eqref{eq:TrSurfTX_N}.
This expression provides a suitable approximation for the CR of an absorbing RX and a matrix TX for $\ratio \ll 1$. The corresponding approximation error is further investigated in Appendix~\ref{sec:ap_up_chDom}. 
%We note that an identical expression for \eqref{eq:cr:beta-} can be derived from the alternative formulation of $N(t)$ on the right hand side of \eqref{eq:cr:conv1}, i.e., $\tilde{N}(t)\big\vert_{\ratio \ll 1} = M_\infty \delta(t) \ast N_\mathrm{s}(t)$.

\subsection{Release Dominated Regime}

In the \textit{release dominated regime}, i.e., $\trel \gg \tabs$ and $\ratio \gg 1$, the molecule release process determines the characteristics of the overall CR, while the exact channel dynamics have less influence.
Inspecting \eqref{eq:ratio:2}, this regime is reached for, e.g., small distances $d$ between TX and RX, and for a slow release of molecules from the TX due to a small $\Dm$ or large $\nacs$ for matrix type TXs.
%Similar to \eqref{eq:cr:beta-}, we propose an approximation for the overall CR which exploits that the CR is dominated by the dynamics of the release process, while the influence of the impulse response is less important.
%
In this regime, the exact dynamics of the channel do not contribute to the overall CR in \eqref{eq:cr:conv1}, and thus, we approximate the  CR as follows 
\begin{align}
	N(t) = M(t) \ast p_\mathrm{s}(t) \approx  
	\int_{t - \tabs}^{t}M(\xi)p_\mathrm{s}(t - \xi)\dint{\xi} = \tilde{N}(t)\big\vert_{\ratio \gg 1}.
	\label{eq:cr:beta+:1}
\end{align}
Since $\tabs \ll \trel$, the value of $M(\xi)$ is almost constant for $t - \tabs < \xi < t$ and can be extracted from the convolution integral
\begin{align}
	\tilde{N}(t)\big\vert_{\ratio \gg 1} = M(t) \int_{t - \tabs}^{t} p_\mathrm{s}(t - \xi)\dint{\xi} = M(t) \int_{0}^{\tabs} p_\mathrm{s}(\xi')\dint{\xi'},
	\label{eq:cr:beta+:2}
\end{align}
where the integral on the right hand side is obtained by the substitution $\xi' = t - \xi$. 
Since the hitting probability $p_\mathrm{s}$ tends to zero for $t \gg \tabs$, a final approximation for the overall CR in the \textit{release dominated regime} is proposed as follows (returning from $\xi'$ to $\xi$)
\begin{align}
	\tilde{N}(t)\big\vert_{\ratio \gg 1} = M(t) \int_0^\infty p_\mathrm{s}(\xi)\dint{\xi} = 
	M(t)\frac{\rrx}{d}
	,
	\label{eq:cr:beta+:3}
\end{align}
where $\frac{\rrx}{d}$ is the final value of  $N_\mathrm{s}$ in \eqref{TS_TX_Abs_RX,Rate}, i.e., $\int_0^\infty p_\mathrm{s}(\xi)\dint{\xi} = \lim_{t\to\infty}N_\mathrm{s}(t) = \frac{\rrx}{d}$.
The closed-form expression in \eqref{eq:cr:beta+:3} provides a suitable approximation for the CR of an absorbing RX and a matrix TX for $\ratio \gg 1$. The resulting approximation error is further investigated in Appendix~\ref{sec:ap_up_reDom}.

\subsection{Discussion}
\vspace*{-1ex}
Approximations \eqref{eq:cr:beta-} and \eqref{eq:cr:beta+:3} for the CR in \eqref{eq:cr:conv1} are applicable when the dynamics of the channel and that of the release process determine the characteristics of the CR, respectively. 
Compared to \eqref{eq:cr:conv1}, which can only be solved numerically except for the special case of $\nacs = 1$, both approximations provide a closed-form expression for the CR. 
In particular, if $\ratio \ll 1$ and $\ratio \gg 1$, \eqref{eq:cr:beta-} and \eqref{eq:cr:beta+:3} can be applied, while the convolution integral has to be solved in the intermediate range. 
%To prove the validity of these approximations, we introduced upper bounds for the approximation error in Appendices~\ref{sec:ap_up_chDom} and \ref{sec:ap_up_reDom}, which are evaluated further in Section~\ref{sec:eval}. 

The approximation for the \textit{channel dominated regime} ($\ratio\ll 1$) in \eqref{eq:cr:beta-} exploits that the impact of the dynamics of the molecule release process is negligible. 
In the limit, the gradual molecule release is replaced by an instantaneous release of molecules from the TX surface.
%, i.e., the proposed approximation is equivalent to the well known point TX assumption. 
This assumption is very common in the MC literature, even if it is not very realistic, and is mostly made because it leads to simpler analytical results.
Instead, approximation \eqref{eq:cr:beta-} shows that the simplification of a gradual release process to an instantaneous release of molecules is valid for $\tau \ll 1$.

%the actual CR turns into an instantaneous release for a given TX-RX arrangement. 
%for, e.g., large distances $d$ between TX and RX, small channel diffusion $\Dc$, or for very small $\nacs$ values in the case of a matrix TX.

The approximation for the \textit{release dominated regime} ($\ratio \gg 1$) in \eqref{eq:cr:beta+:3} exploits that the impact the of dynamics of the channel between TX and RX is negligible. 
Hence, we assume the instantaneous arrival of a molecule at the RX after its release from the TX with hitting probability $\frac{\rrx}{d}$. 
%This approximation is of particular interest if distance $d$ between TX and RX is small and for a slow release of molecules from the TX originating from small $\Dm$ or large $\nacs$-values for matrix type TXs. 
In Section~\ref{subsec:release_micelle}, we observed that the release process of different types of drug molecules from diblock copolymer micelle type drug carriers is rather slow as the corresponding $\nacs$-values are large and the diffusion inside the polymer matrix is very slow.   
The resulting long durations $\trel$ of the molecule release process from practical drug carriers highlight the relevance of the proposed approximation for $\ratio \gg 1$, as it provides a closed-form expression for the CR of realistic DD scenarios.

The characterization of the molecule transmission by the ratio $\tau$ of $\trel$ and $\tabs$ has been derived for the considered scenario of a matrix TX and an absorbing RX in free space. 
However, the ratio $\tau$ can also be applied to characterize other types of MC systems, once $\trel$ and $\tabs$ for the specific system considered have been derived. 
As the approximations \eqref{eq:cr:beta-} and \eqref{eq:cr:beta+:3} are independent from the exact form of $\trel$ and $\tabs$, they can also be applied.

%for the limiting regimes, can also be applied.
%It is important to note that this characterization of the molecule transport is generally not restricted to the considered scenario of matrix type TXs. 
%Independent from the exact type of molecule release, the ratio $\ratio$ between $\trel$ and $\tabs$ provides information whether the release process or the channel dynamics are dominant. 

%% file: num_eval.tex
\begin{table*}[t]
\caption{\small Parameters for numerical evaluation}
\label{tab:parameter}
\centering
\begin{tabular}{p{2.3cm}p{2.3cm}p{2cm}p{1.5cm}p{1cm}p{1cm}p{1.9cm}p{1.2cm}}
\hline\noalign{\smallskip}
&$\Dm$& $\Dc$& $d$ & $a$ &$\rrx$ & $\nacs$ & Refs.\\
\noalign{\smallskip}\hline\noalign{\smallskip}
Secs.~\ref{subsec:num_eval:release}, \ref{subsec:num_eval:CR}  & $10^{-9}\,\si{\square\meter/\second}$ &$10^{-9}\,\si{\square\meter/\second}$& $\{2,\,5\}\,\si{\micro\meter}$ &$1\,\si{\micro\meter}$ & $1\,\si{\micro\meter}$ & $\{1,\,25,\,100,\,400\}$ &\footnotesize\cite{noel:16} 	\smallskip\\
Sec.~\ref{subsec:validRegimes}  & $10^{-8}\,\si{\square\meter/\second}$ &$10^{-8}\,\si{\square\meter/\second}$& $20\,\si{\micro\meter}$ &$1\,\si{\micro\meter}$ & $1\,\si{\micro\meter}$ & $\{1,\dots,10^{11}\}$ & \smallskip\\
\noalign{\smallskip}\hline\noalign{\smallskip}
Sec.~\ref{subsec:eval:param}  &  &&  && &  \smallskip\\
DOX pH~$5.0$& $1.82\cdot 10^{-22}\,\si{\square\meter/\second}$ &$5\cdot 10^{-11}\,\si{\square\meter/\second}$& $10\,\si{\micro\meter}$ & $4.5\,\si{\nano\meter}$ &$5\,\si{\micro\meter}$ & $63.7$ &\hspace*{-0.8cm}\footnotesize\cite{Sutton2007,Weinberg2007,SUN2000184} \\
DOX pH~$7.4$& $1.82\cdot 10^{-22}\,\si{\square\meter/\second}$ &$5\cdot 10^{-11}\,\si{\square\meter/\second}$& $10\,\si{\micro\meter}$ & $4.5\,\si{\nano\meter}$ &$5\,\si{\micro\meter}$ & $757.5$ &\hspace*{-0.8cm}\footnotesize\cite{Sutton2007,Weinberg2007,SUN2000184} \\
$\beta$-lap& $2.42\cdot 10^{-21}\,\si{\square\meter/\second}$&  $1\cdot 10^{-9}\,\si{\square\meter/\second}$ & $10\,\si{\micro\meter}$& $4.5\,\si{\nano\meter}$ &$5\,\si{\micro\meter}$ & $370.4$ & \hspace*{-0.8cm}\footnotesize\cite{Sutton2007, SUN2000184} \\
\noalign{\smallskip}\hline\noalign{\smallskip}
\end{tabular}
\vspace*{-6ex}
\end{table*}

In the following, we numerically evaluate the influence of the considered practical matrix TX model on the CR of a diffusive MC system with an absorbing RX (see Fig.~\ref{fig:system}). 
First, we investigate the molecule release process for different $\nacs$-values from a matrix TX in Section~\ref{subsec:num_eval:release}. 
In Section~\ref{subsec:num_eval:CR}, we compare the CR of the matrix TX with those of a point TX and a transparent spherical TX~\cite{noel:16}. 
The validity of the proposed approximations for the limiting regimes introduced in Section~\ref{sec:regimes} are investigated in Section~\ref{subsec:validRegimes}. 
Finally, we study a practical DD scenario in Section~\ref{subsec:eval:param}, where we consider a diblock copolymer micelle matrix TX and two types of drug molecules, as discussed in Section~\ref{subsec:release_micelle}.

For the evaluation, we adopt the parameter values in Table~\ref{tab:parameter}.
Initially, the different TX types are loaded with $M_\infty = 10^4$ molecules, where the molecules are uniformly distributed over the entire volume or are concentrated in a point. 
 At $t=0$, for the point TX and the spherical TX, the molecules are instantaneously released, and for the matrix TX, the molecules are gradually released. 
For validation of the expressions derived for the CR of the matrix TX, we embedded the proposed PBS model for the matrix system~(see~Section~\ref{sec:drug_pbs}) in the PBS model of a diffusive MC system with an absorbing RX. 
For the PBS results in Sections~\ref{subsec:num_eval:release} and \ref{subsec:num_eval:CR}, the refined Monte Carlo (RMC) algorithm was used with a time step of~$\Delta t = 1\cdot 10^{-6}\,\si{\second}$ \cite{Arifler:ieee:2017}.  
Due to the large simulation durations for the practical release processes in Section~\ref{subsec:release_micelle}, the a priori Monte Carlo (APMC) algorithm was used with a time step of $\Delta t = 360\,\si{\second}$ \cite{Wang:ieee:2018}. 
All PBS results were averaged over $100$ realizations.

%All PBS results were obtained with a time step of~$\Delta t = 1\cdot 10^{-6}\,\si{\second}$ and were averaged over $100$ realizations.

\subsection{Molecule Release Process}
\label{subsec:num_eval:release}
\vspace*{-1ex}

\begin{figure*}
	\centering
	\begin{subfigure}[b]{0.49\linewidth}
            \centering
            	\includegraphics[width=\linewidth]{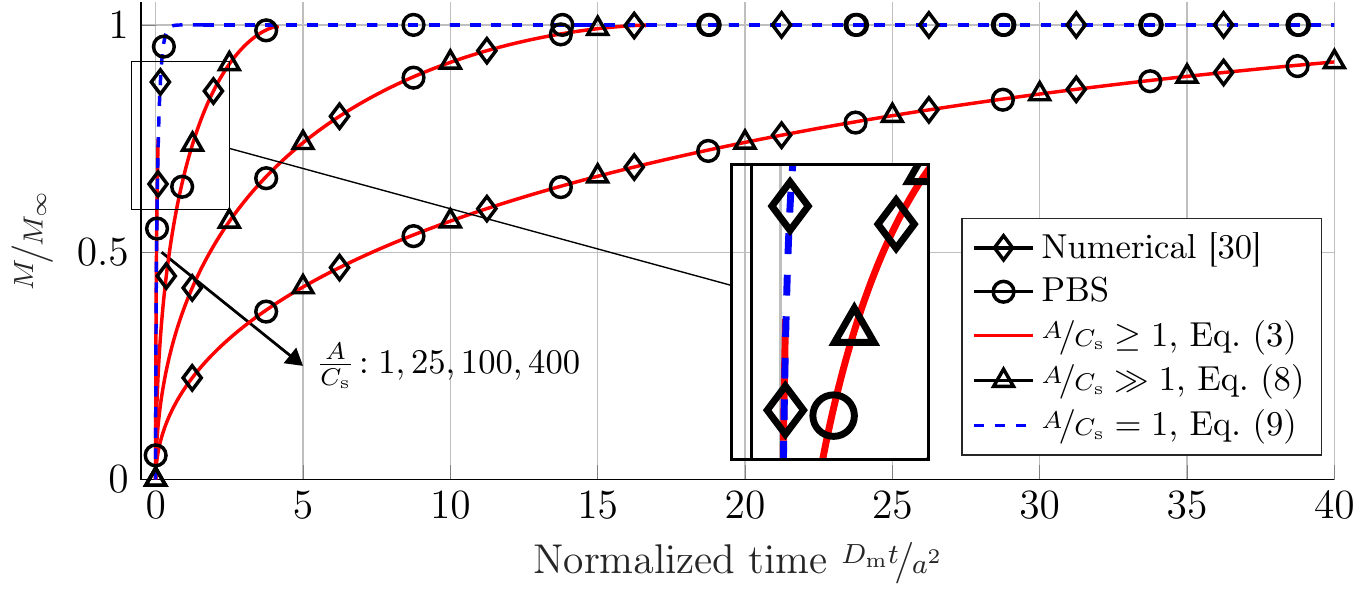}
            	\vspace*{-5ex}
            \caption{Normalized molecule release}
    \label{fig:profile:m}
    \end{subfigure}
    \begin{subfigure}[b]{0.49\linewidth}
            \centering
            	\includegraphics[width=\linewidth]{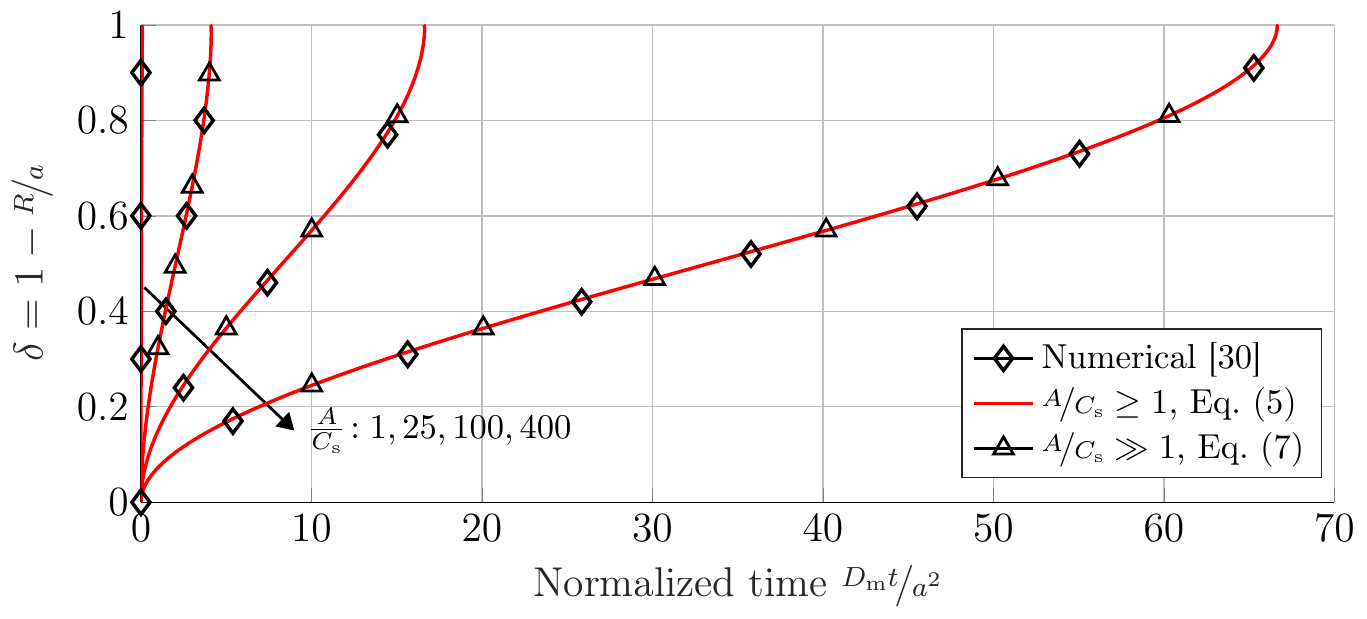}
            	\vspace*{-5ex}
            \caption{Diffusion front position}
    \label{fig:profile:del}
    \end{subfigure}
    \vspace*{-2ex}
    \caption{\small (a) Normalized release $M(t)/M_\infty$ over normalized time $\nicefrac{\Dm t}{a^2}$, and (b) corresponding time-dependent position of diffusion front $\delta$. The FDM solution according to \cite{Koizumi:1995}, the results from PBS according to Section~\ref{sec:drug_pbs}, the approximate solutions in \eqref{eq:9}, \eqref{eq:10} and \eqref{eq:frenning:delta}, \eqref{eq:frenning:M}, and the solution for $\nacs = 1$ in \eqref{eq:crankVinf1} are shown.}
    \label{fig:profile}
    \vspace*{-5ex}
\end{figure*}

First, we compare the models for the release of molecules from a matrix TX as described in Section~\ref{sec:drug}. 
In Fig.~\ref{fig:profile:m}, the normalized number of released molecules $M/M_\infty$ is plotted over normalized time $\nicefrac{\Dm t}{a^2}$ for different $\nacs$-values. 
We observe that the amount of molecules released until a certain time decreases with increasing $\nacs$, i.e., the release rate reduces for increasing~$\nacs$. 
Fig.~\ref{fig:profile:del} shows the corresponding time-dependent position of the diffusion front $\delta$ derived from \eqref{eq:10} and \eqref{eq:frenning:delta}, respectively, and reveals that the speed of the diffusion front decreases with increasing $\nacs$.
From both subfigures of Fig.~\ref{fig:profile}, we observe that the exact FDM solution from \cite{Koizumi:1995} (diamond markers), the results from the proposed PBS according to Section~\ref{sec:drug_pbs} (circle markers), and the approximate solutions \eqref{eq:9}, \eqref{eq:10} (red curves) and \eqref{eq:frenning:M}, \eqref{eq:frenning:delta} (triangle markers) are in excellent agreement for all considered $\nacs > 1$. 
%Also the simplified model in \eqref{eq:frenning:M} that is a direct function of time is in excellent agreement with the exact numerical solution. 

However, for $\nacs = 1$, we observe that the approximate solution \eqref{eq:9} is not applicable as can be seen from the zoomed area in Fig.~\ref{fig:profile:m}. 
In this case, the approximate solution does not reach $1$, i.e., $\frac{M(1)}{M_\infty} = 1 - \frac{C_\mathrm{s}}{4A} = 0.75$, and is not able to capture the complete dynamics of the release process. 
Nevertheless, as $\nacs = 1$ corresponds to an instantaneous release from a transparent sphere, \eqref{eq:crankVinf1} can be applied instead of \eqref{eq:9} and the corresponding results are in excellent agreement with the numerical results, as shown in Fig.~\ref{fig:profile:m} (dashed blue line).

\subsection{Channel Response}
\label{subsec:num_eval:CR}
\vspace*{-1ex}

\begin{figure*}[!t]
	\begin{minipage}[t]{0.49\linewidth}
		\centering
		\includegraphics[width=\linewidth]{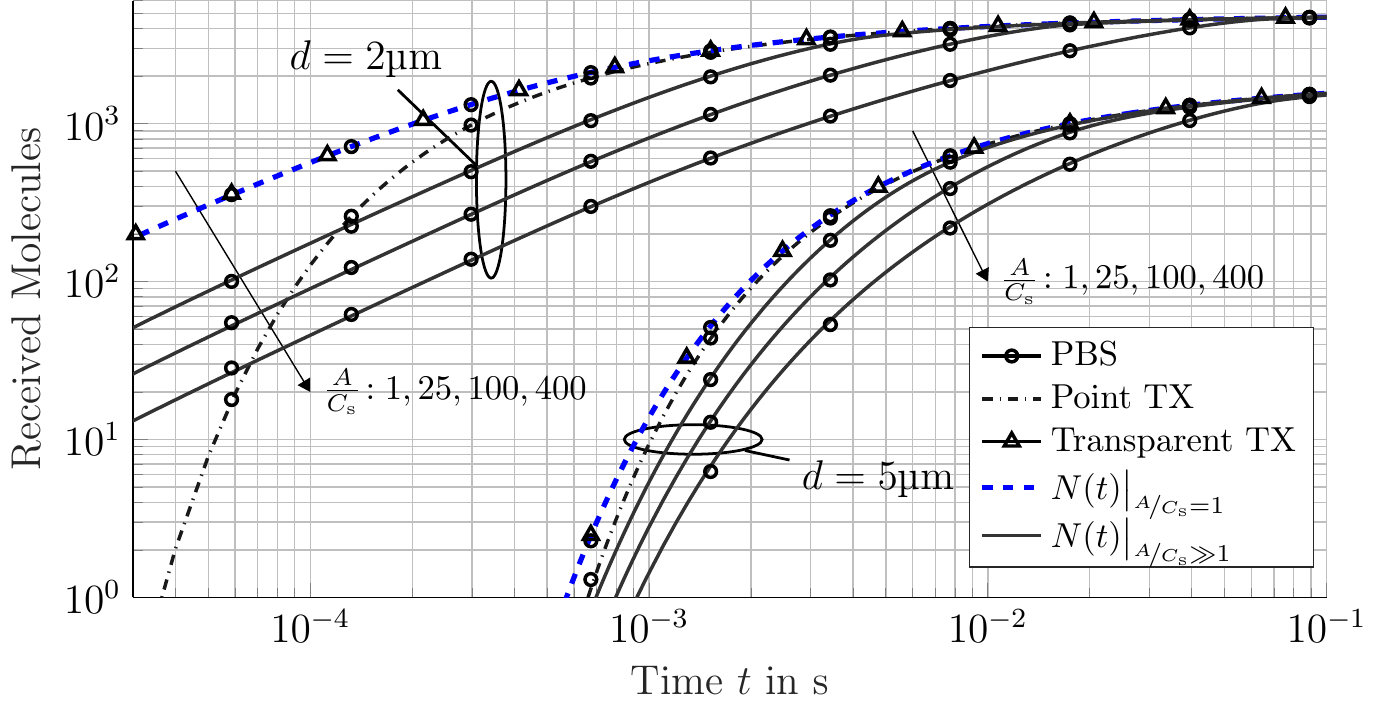} 
            	\vspace*{-6ex} 
 	\caption{\small CR for an absorbing RX due to a point release (dash dotted lines), a spherical release (triangle markers), and matrix release for different $\nicefrac{A}{C_\mathrm{s}}$-values (blue and black lines) for $d \in \{2,5\}\,\si{\micro\meter}$. Results from PBS are shown as circle markers.}
 	\label{fig:3}
	\end{minipage}\hfill
	\begin{minipage}[t]{0.49\linewidth}
		\centering
            	\includegraphics[width=\linewidth]{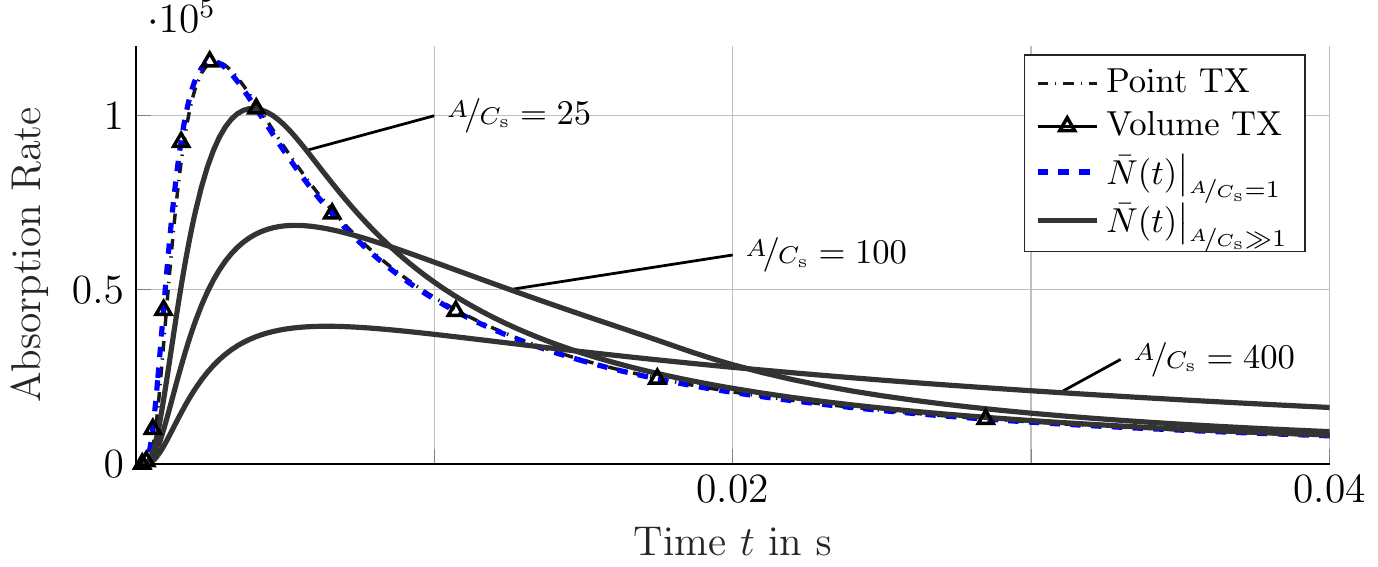}  
            	\vspace*{-6ex}
 	\caption{\small Absorption rate $\bar{N}(t)$ at the RX due to a point release (dash dotted line), a spherical release (triangle markers), and a matrix release for different $\nicefrac{A}{C_\mathrm{s}}$ (blue and black lines) for $d = 5\,\si{\micro\meter}$.}
 	\label{fig:absRate}
 	 \vspace*{-3ex}
	\end{minipage}
	\vspace*{-5ex}
\end{figure*}

In Fig.~\ref{fig:3}, we investigate the CR for a matrix TX and an absorbing RX as derived in Section~\ref{sec:channel}. 
The figure shows the number of molecules received by an absorbing RX for $d = \{2,\,5\}\,\si{\micro\meter}$ and different types of TXs, i.e., a point TX (dash dotted lines), a transparent sphere TX (triangle markers)~\cite{noel:16}, and a matrix TX with $\nacs = 1$ (Eq.~\eqref{eq:response_ins2}, dashed blue line) and $\nacs \gg 1$ (Eq.~\eqref{eq:response_con}, black lines). The results from PBS are shown by circle markers for validation.

First, we observe that the results for the point, transparent, and matrix TX are in excellent agreement with the results from PBS for all considered scenarios. 
Furthermore, the differences between the considered TX types are more pronounced when the distance between TX and RX is smaller. 
The characteristic for all TX types becomes similar as time increases.
For $\nacs = 1$, we observe that the CR in \eqref{eq:response_ins2} is equivalent to the release from a transparent sphere. 
This confirms that \eqref{eq:response_ins2} is an analytical expression for the spherical TX model proposed in \cite{noel:16}.

Moreover, for $d = 2\,\si{\micro\meter}$ we observe that for the point TX, it takes longer for the first molecules to arrive at the RX compared to the spherical and matrix TXs. 
This is because all molecules are initially located at the center of the TX and not in the outer layers of the sphere. For $d = 5\,\si{\micro\meter}$ this effect is less pronounced. 

For the matrix TX, we observe that the absorption of molecules at the RX slows down for increasing $\nacs$-values.
The reason for this behavior is that for higher $\nacs$-values, the molecules dissolve more slowly from the matrix (see Fig.~\ref{fig:profile:m}). 
For larger $\nacs$, we observe that the delivery of molecules by the matrix TX is more spread over time compared to the point and spherical TX.   
This behavior is desirable for DD systems, where the amount of drugs absorbed by the RX (e.g., a cancer cell) should stay in a desired range during the delivery process \cite[Fig.~1]{Sutradhar2016}.

We investigate this behavior more in detail in Fig.~\ref{fig:absRate}, which shows the absorption rate $\bar{N}(t) = \frac{\mathrm{d}}{\mathrm{d}t}N(t)$ of the RX for different TX types and $d = 5\,\si{\micro\meter}$. 
We observe that the absorption rates for the point, transparent, and matrix TX ($\nacs = 1$) are nearly identical. 
In particular, the absorption rate increases very fast for $t >0$, but decreases again very fast as the molecule release at the TX was instantaneous. 
For the gradual release of molecules by the matrix TX ($\nacs > 1$), we observe that the absorption rate is spread over time because the release time \eqref{eq:release_time} of the matrix TX scales with $\nacs$. 
This reveals the importance of parameter $\nacs$ for the design of DD systems with practical carriers.
In particular, the amount of delivered drugs over time can be controlled by $\nacs$, i.e., by loading $A$ and drug solubility $C_\mathrm{s}$. 
%
% i.e., by a specific variation of loading $A$ or solubility $C_\mathrm{s}$ the amount of delivered drugs over time can be controlled.

\subsection{Validation of Limiting Regimes}
\label{subsec:validRegimes}
\vspace*{-1ex}

In the following, we evaluate the validity of the approximations for the \textit{channel dominated} and \textit{release dominated regimes} proposed in Section~\ref{sec:regimes}. 
We employ the Normalized Root Mean Square Error (NRMSE) between the approximated and actual CRs, as well as the $\%$-deviations of both CRs. The NRMSE of the difference $\Delta_\mathrm{CR}(t)$ between the approximate value $\tilde{N}(t)$ and the actual value $N(t)$ can be defined as follows \cite{Li:rmse:2006}
\begin{align}
	\mathrm{NRMSE}\{\Delta_\mathrm{CR}\} = \frac{1}{N(t\to\infty)} 
	\left(\int_0^\infty \Delta_\mathrm{CR}^2(\xi)\dint{\xi} \right)^{\frac{1}{2}}
%	\sqrt{\mathrm{E}\left[ \tilde{\bm{x}}^2\right]}= \sqrt{\mathrm{E}\left[ (\hat{\bm{x}} - \bm{x})^2\right]},
	\label{eq:rmse}
\end{align}
%where $N_\infty$ is the final value of the actual CR, i.e., $N_\infty = \lim_{t\to\infty}N(t)$.
The parameters for numerical evaluation are given in Table~\ref{tab:parameter} and, for the calculation of $\tabs(\sigma)$, we used $\sigma = 0.99$, i.e., $\tabs$ is the time until $99\%$ of $M_\infty$ particles are absorbed by the RX.  
From a practical point of view, the properties of the TX are the most important tuning parameters.
Therefore, the variation of the regime, and consequently $\ratio$ in \eqref{eq:ratio:2}, is achieved by a variation of the $\nacs$-values, while all other parameters remain constant\footnote{We note that the other parameters in \eqref{eq:ratio:2} may also be varied to achieve different regimes, e.g., a variation of $\Dm$ via different polymer structures.}.
%

%where $\tilde{\bm{x}}$ is the difference between an approximate value $\hat{\bm{x}}$ and the actual value $\bm{x}$, i.e., $\tilde{\bm{x}} = \hat{\bm{x}} - \bm{x}$.
%The error between the approximations and the actual CR will tend to zero for increasing time, i.e., $\Delta_{\mathrm{CR}}\to 0$ for $t \to \infty$\footnote{As both approximations can be interpreted as the simplification of the TX model, the reason is that for $t\to\infty$ all TX types yield the same CR, c.f., \cite{noel:16}.}. Hence, we have chosen the RMSE as an appropriate error measure, as the RMSE tends to privilege large errors, occurring at the beginning of the CR, compared to other error metrics \cite{Li:rmse:2006}. 

\begin{figure*}
	\centering
	\includegraphics[width=0.7\linewidth]{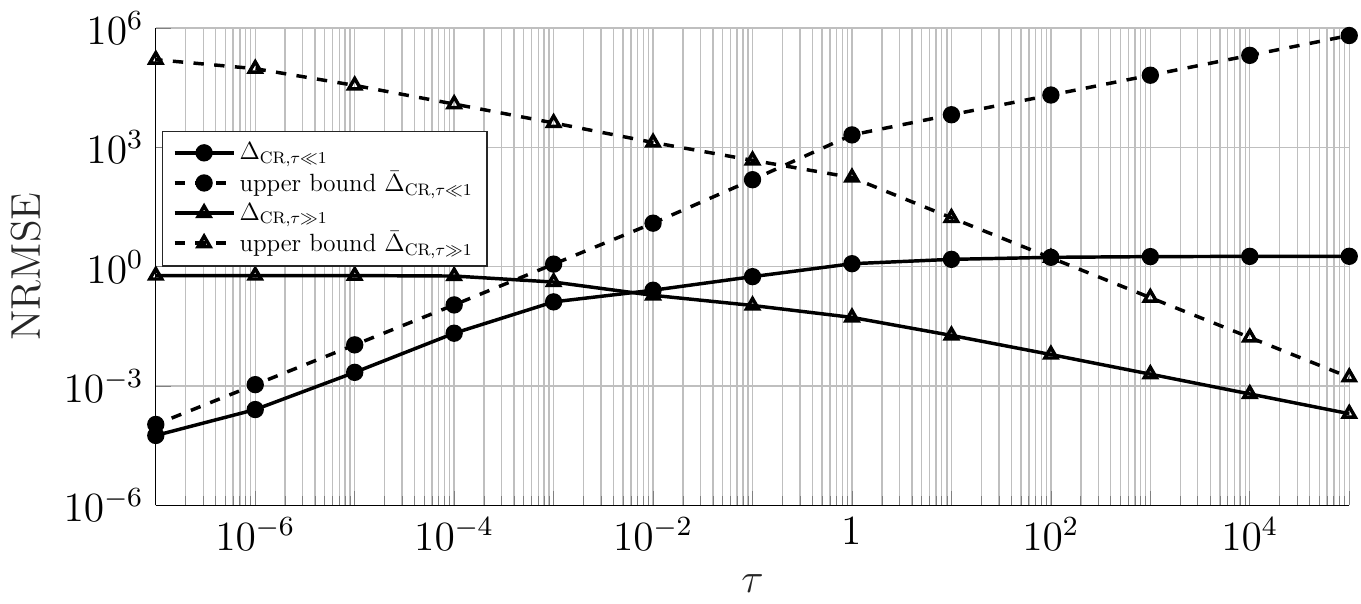}
	\vspace*{-2ex}
	\caption{\small NRMSE of the approximations for the channel dominated and release dominated regimes (solid lines) as well as the corresponding upper error bounds (dashed lines). The RMSE is calculated by \eqref{eq:rmse}, with the approximation errors $\Delta_{\mathrm{CR}, \ratio\gg 1}$ in \eqref{eq:cr:beta+:app:1} and $\Delta_{\mathrm{CR}, \ratio\ll 1}$ \eqref{eq:cr:beta-:app:1}, and upper bounds $\bar{\Delta}_{\mathrm{CR}, \ratio\gg 1}$ in \eqref{eq:cr:beta+:app:5} and $\bar{\Delta}_{\mathrm{CR}, \ratio\ll 1}$ in \eqref{eq:cr:beta-:app:5}.}
	\label{fig:beta:rmse}
	\vspace*{-5ex}
\end{figure*}

Fig.~\ref{fig:beta:rmse} shows the NRMSEs of the approximation errors $\Delta_{\mathrm{CR},\ratio \ll 1}$ of $\tilde{N}\vert_{\ratio\ll 1}$ and $\Delta_{\mathrm{CR},\ratio \gg 1}$ of $\tilde{N}\vert_{\ratio\gg 1}$ in \eqref{eq:cr:beta-:app:1} and \eqref{eq:cr:beta+:app:1}, respectively.
%
%Fig.~\ref{fig:beta:rmse} shows the RMSE between the approximations $\tilde{N}\vert_{\ratio\ll 1}$ and $\tilde{N}\vert_{\ratio\gg 1}$ from \eqref{eq:cr:beta-} and \eqref{eq:cr:beta+:3}, respectively, and the actual CR for $\nacs \gg 1$ in \eqref{eq:response_con}. 
%The RMSEs of $\Delta_{\mathrm{CR},\ratio \gg 1}$ in \eqref{eq:cr:beta+:app:1} and $\Delta_{\mathrm{CR},\ratio \ll 1}$ in \eqref{eq:cr:beta-:app:1} are calculated by \eqref{eq:rmse}. 
Furthermore, the NRMSEs of $\bar{\Delta}_{\mathrm{CR},\ratio \ll 1}$ and $\bar{\Delta}_{\mathrm{CR},\ratio \gg 1}$ in \eqref{eq:cr:beta-:app:5} and \eqref{eq:cr:beta+:app:5} are shown as upper bounds for the approximation errors.
%To provide a meaningful representation, the RMSE values are normalized to the actual value of the actual CR.
%
First, we observe from Fig.~\ref{fig:beta:rmse} that the NRMSE for the approximation $\tilde{N}\vert_{\ratio\ll 1}$ in the \textit{channel dominated regime} decreases for decreasing ratios $\ratio$ (solid line with circle markers). 
Second, we observe that the proposed upper bound $\bar{\Delta}_{\mathrm{CR},\ratio \ll 1}$ (dashed line with circle markers) for the approximation always lies above the actual approximation error and becomes tighter with decreasing $\ratio$. 
We observe a similar behavior for the NRMSE of the approximation $\tilde{N}\vert_{\ratio\gg 1}$ in the \textit{release dominated regime} (solid line with triangle markers). 
For increasing values of $\ratio$, the approximation error decreases significantly. 
Furthermore, we observe that the proposed upper bound $\bar{\Delta}_{\mathrm{CR},\ratio \gg 1}$ (dashed line with triangle markers) is more conservative than for the \textit{channel dominated regime}, but also becomes tighter for increasing $\ratio$. 
The results in Fig.~\ref{fig:beta:rmse} confirm that the approximations proposed in Section~\ref{sec:regimes} are valid when either the channel or the release process determine the characteristic of the CR.
In particular, for $\ratio \ll 1$ and $\ratio \gg 1$, the closed-form approximations \eqref{eq:cr:beta-} and \eqref{eq:cr:beta+:3} can be applied instead of the actual CR in \eqref{eq:response_con} which has to be evaluated numerically.
%
%, the closed-form CR in \eqref{eq:cr:beta-} can be applied instead of the actual CR in \eqref{eq:response_con} which has to be evaluated numerically. 
%Similarly, when $\ratio \gg 1$, the closed-form CR in \eqref{eq:cr:beta+:3} can be applied instead of \eqref{eq:response_con}. 

\begin{figure*}
	\centering
	\begin{subfigure}[b]{0.49\linewidth}
            \centering
            \includegraphics[width=\linewidth]{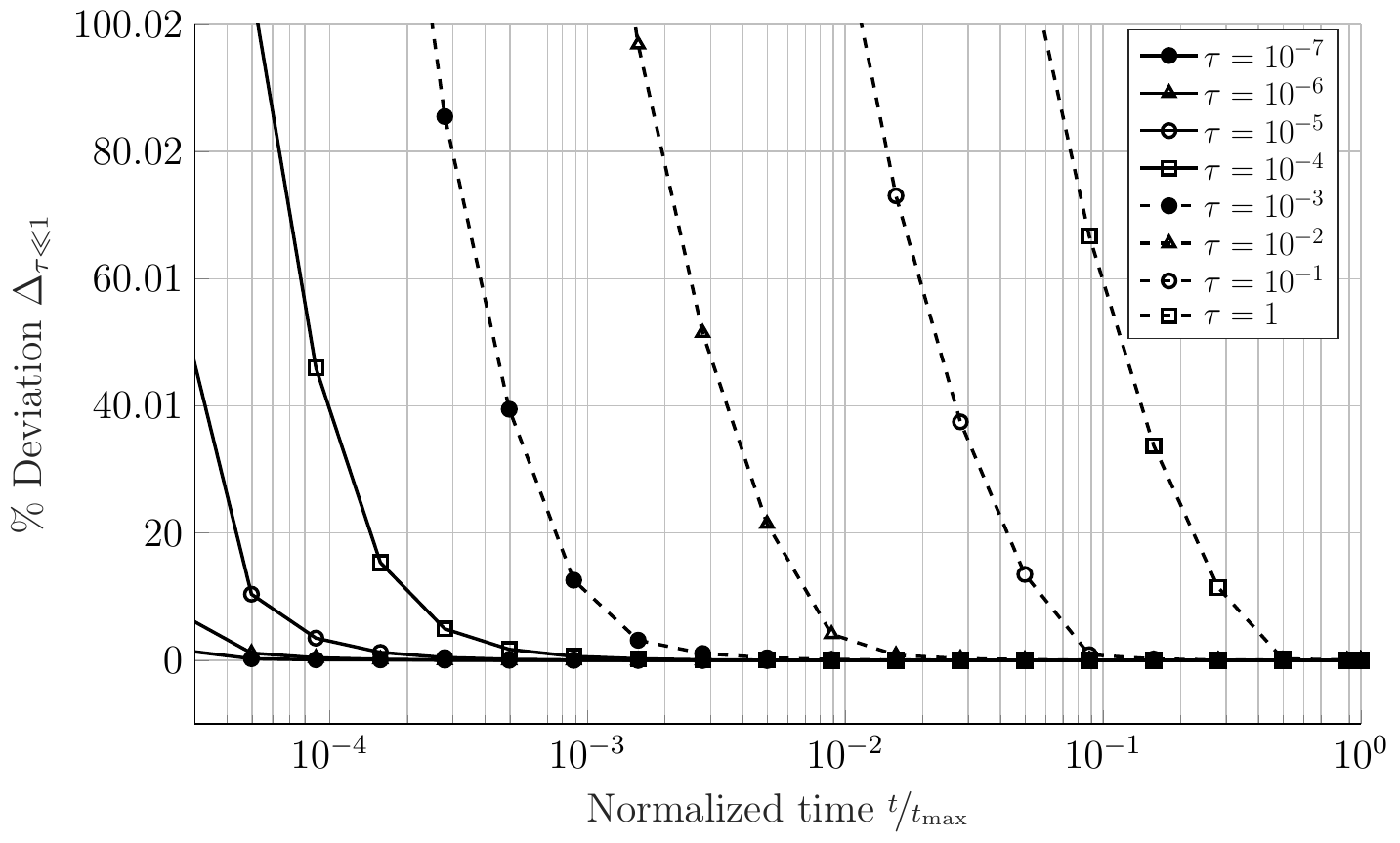}
            	\vspace*{-5ex}
            \caption{Channel dominated regime }
    \label{fig:beta:deviation:instRel}
    \end{subfigure}
    \begin{subfigure}[b]{0.49\linewidth}
            \centering
            	\includegraphics[width=\linewidth]{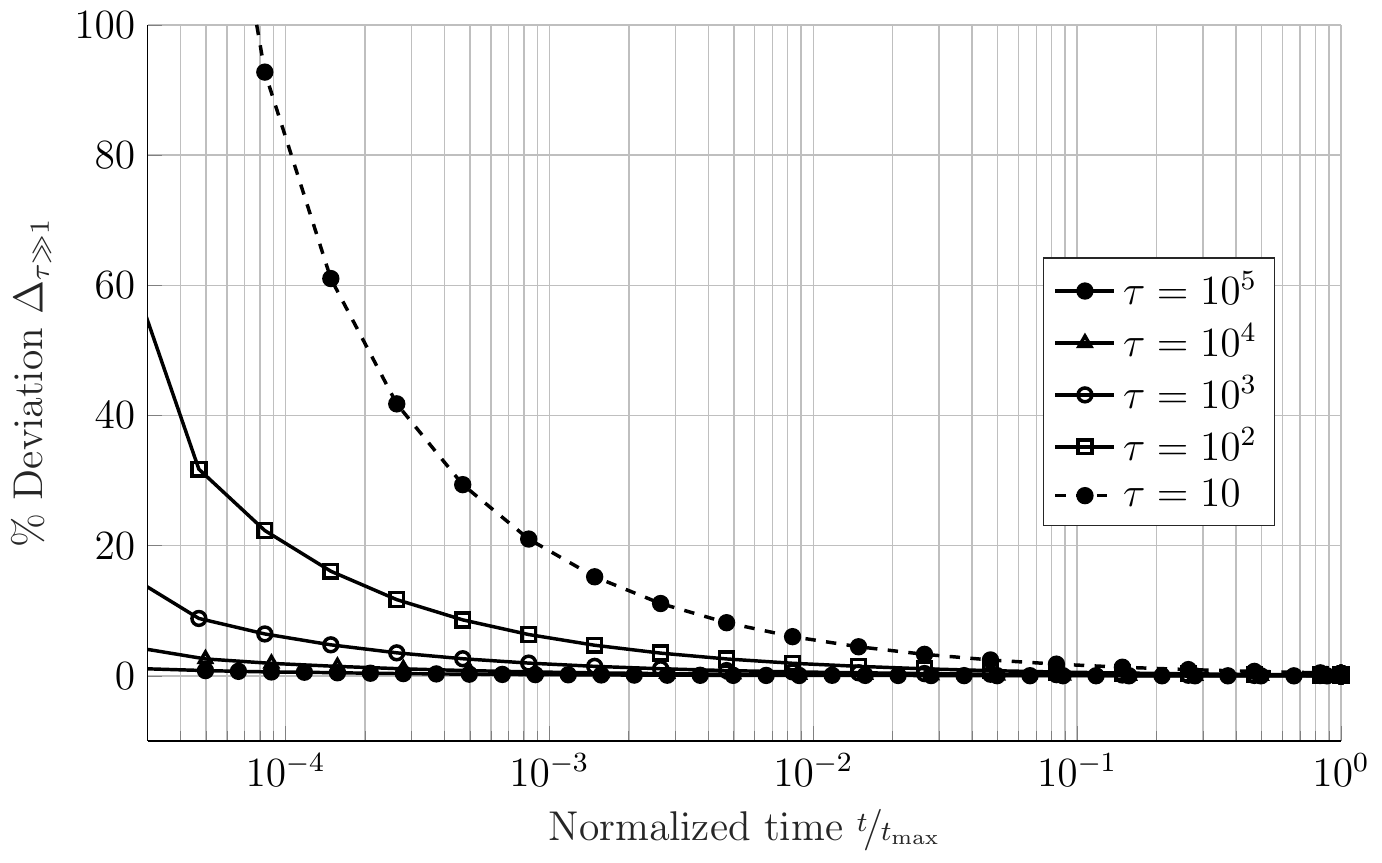}
            \vspace*{-5ex}
            \caption{Release dominated regime}
    \label{fig:beta:deviation:instArriv}
    \end{subfigure}
    \vspace*{-2ex}
    \caption{\small $\%$ deviation \eqref{eq:beta:deviation} of the proposed approximations for the (a) channel dominated regime according to \eqref{eq:cr:beta-} and the (b) release dominated regime according to \eqref{eq:cr:beta+:3}, for different values of the ratio $\ratio$ (different line styles). The arrival times are normalized to $\tmax$ where almost all molecules are absorbed by the RX.
    }
    \label{fig:beta:deviation}
    \vspace*{-5ex}
\end{figure*}

To investigate further for which $\ratio$-values the proposed CRs provide a sufficiently accurate approximation for the actual CRs, the $\%$-deviation is considered, i.e.,
\begin{align}
	&\Delta_{\ratio \ll 1} = \frac{\Delta_{\mathrm{CR}, \ratio \ll 1}}{N(t)}\times 100\%,
%	&\Delta_{\ratio \ll 1} = \left({\Delta_{\mathrm{CR}, \ratio \ll 1}}/{N(t)}\right)\times 100\%,	 
	&\Delta_{\ratio \gg 1} = \frac{\Delta_{\mathrm{CR}, \ratio \gg 1}}{N(t)}\times 100\%.
	\label{eq:beta:deviation}
\end{align} 
The corresponding results are shown in Fig.~\ref{fig:beta:deviation}. 
To unify the duration of the considered CRs, the time scale in Figs.~\ref{fig:beta:deviation:instRel} and \ref{fig:beta:deviation:instArriv} is normalized to $\tmax = \tabs + \trel$, where almost all molecules are absorbed by the RX, i.e., 
$N(t = \tmax) \approx N(t\to \infty)$. 
%The value of $\tmax$ can be derived from $N(\tmax)$ and the maximum of $\tabs$ and $\trel$, i.e., $\tmax = 4\mathrm{max}(\trel, \tabs)$. 
In Fig.~\ref{fig:beta:deviation:instRel}, the $\%$-deviation between the approximated CR \eqref{eq:cr:beta-} and the actual CR \eqref{eq:response_con} is shown for different values of ratio $\ratio \leq 1$. 
First of all, we observe that the deviation tends to zero for $\nicefrac{t}{\tmax}\to 1$.
%, which reconfirms the RMSE as an appropriate error measure. 
Furthermore, the curves support the observations in Fig.~\ref{fig:beta:rmse}, i.e., the deviation decreases with decreasing $\ratio$. 
For $\ratio= 10^{-5}$, the deviation drops to $0\%$ at $\nicefrac{t}{\tmax} \approx 4\cdot 10^{-4}$ ($t = 36\,\si{\milli\second}$). The corresponding NRMSE value for $\ratio = 10^{-5}$ is $10^{-3}$ and it decreases further for $\tau < 10^{-5}$ (see Fig.~\ref{fig:beta:rmse}). 
%Hence, we consider that for the \textit{channel dominated regime} \eqref{eq:cr:beta-} is an accurate approximation for $\ratio \leq 10^{-5}$.

Fig.~\ref{fig:beta:deviation:instArriv} shows the $\%$-deviation between the approximated CR \eqref{eq:cr:beta+:3} and actual CR \eqref{eq:response_con} for different values of ratio $\ratio \geq 1$. 
In contrast to the case of $\ratio \leq 1$, we observe that for $\tau \geq 1$ the deviation decreases for increasing values of $\ratio$. 
For $\ratio = 10^3$, the deviation drops to $0\%$ for $\nicefrac{t}{\tmax} \approx 4\cdot 10^{-3}$ ($t = 300\,\si{\second}$). 
The corresponding NRMSE value for $\ratio = 10^3$ is $10^{-3}$ and it decreases further for $\tau > 10^3$ (see Fig.~\ref{fig:beta:rmse}). 
%Therefore, the approximation \eqref{eq:cr:beta+:3} for the \textit{release dominated regime} is accurate for $\ratio \geq 10^{3}$.

The above analysis confirms that approximations \eqref{eq:cr:beta-} and \eqref{eq:cr:beta+:3} for the actual CR \eqref{eq:response_con} are accurate for the limiting regimes, when either the channel dynamic or the molecule release dynamic determine the characteristics of the CR. 
%For $\ratio < 10^{-5}$, the proposed CR \eqref{eq:cr:beta-} provides a sufficient approximation for the actual CR by assuming an instantaneous release of molecules instead of the exact TX dynamics. 
%
%This goes along with the common point TX assumption that is often made in the study of MC system. 
%For $\ratio > 10^{3}$, the proposed CR \eqref{eq:cr:beta+:3} provides a sufficient estimate for the actual CR, by assuming an instantaneous arrival of all molecules, which results in a CR that is a weighted version of the molecule release process.
%
%Thus the proposed CR \eqref{eq:cr:beta+:3} for the \textit{release dominating regime} provides a sufficient estimate for the actual CR for $\ratio > 10^{3}$.
%
%the actual CR can be simplified by the assumption of an instantaneous arrival of all molecules, which results in an CR that is a weighted version of the molecule release process. 
%
%
Furthermore, the analysis showed that the \textit{release dominated regime} is of particular interest for the investigation of practical DD systems. 
As has been discussed in Section~\ref{subsec:release_micelle}, for practical carrier dimensions and drug molecules, the $\nacs$-values are rather large and the diffusion coefficient $\Dm$ inside the polymer is very small. 
According to \eqref{eq:release_time} and \eqref{eq:ratio:2}, these values lead to large release durations and large $\ratio$-values.
The application of the closed-form CR \eqref{eq:cr:beta+:3} for the analysis of practical DD systems is discussed in the following.

\subsection{Drug Release from Diblock Copolymer Micelles}
\label{subsec:eval:param} 
\vspace*{-1ex}
 
 \begin{figure*}
	\centering
	\begin{subfigure}[b]{0.49\linewidth}
            \centering
            	\includegraphics[width=\linewidth]{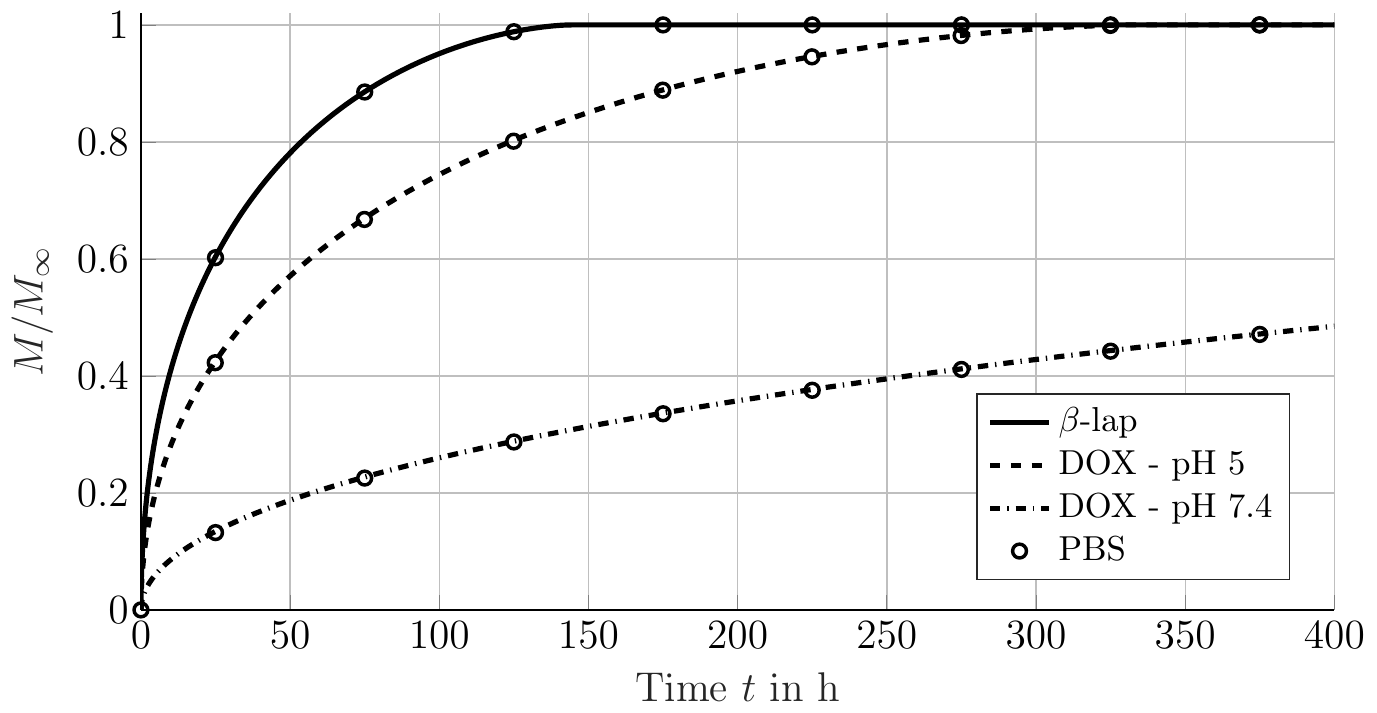}
            	\vspace*{-5ex}
            \caption{Drug Molecule Release}
    \label{fig:real:m}
    \end{subfigure}
    \begin{subfigure}[b]{0.49\linewidth}
            \centering
            	\includegraphics[width=\linewidth]{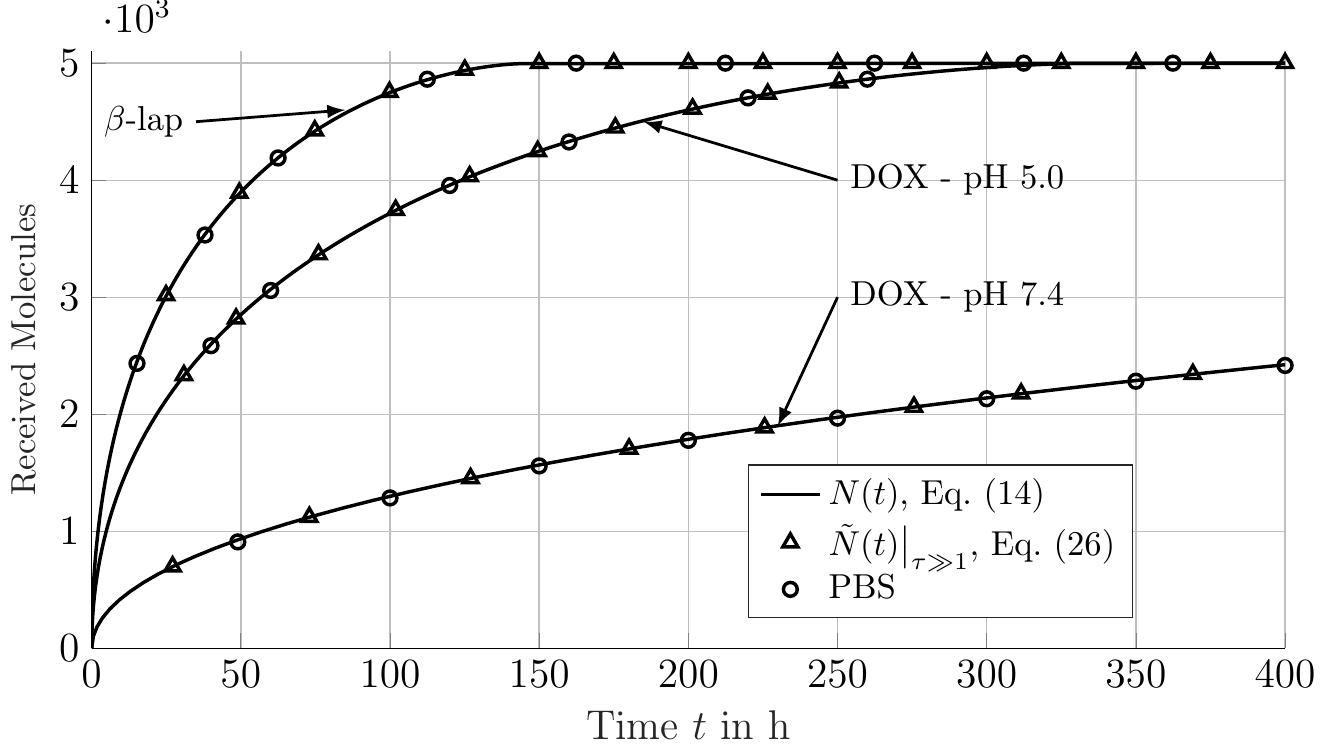}
            	\vspace*{-5ex}
            \caption{Channel Response}
    \label{fig:real:cr}
    \end{subfigure}
    \vspace*{-2ex}
    \caption{\small (a) Normalized amount of released molecules $M(t)/M_\infty$ over normalized time $t$ according to \eqref{eq:frenning:M_v2}, and (b) CR for an absorbing RX and a matrix release of $\beta$-lap and DOX from a diblock copolymer micelle core. Actual impulse response according to \eqref{eq:cr:conv1}, approximate solution for $\ratio \gg 1$ according to \eqref{eq:cr:beta+:3}, and results from PBS. 
    }
    \label{fig:real}
    \vspace*{-4ex}
\end{figure*}
 
In the following, we investigate the CR for a practical DD system, consisting of the diblock copolymer micelle matrix TX (see Section~\ref{subsec:release_micelle}) and an absorbing RX representing, e.g., the nucleus of a diseased cell. 
As drug molecules we consider $\beta$-lap and DOX at two different pH levels. 
The corresponding parameters are summarized in the lower part of Table~\ref{tab:parameter}.

Fig.~\ref{fig:real:m} shows the results from PBS (circle markers) and the results from \eqref{eq:frenning:M_v2} (solid, dashed, and dash dotted lines) for the amount of molecules released from the micelle core of radius $a$ loaded with $\beta$-lap and DOX molecules at different pH levels (see Fig.~\ref{fig:micelle}).
%The individual curves in Fig.~\ref{fig:real:m} are 
%the results from PBS (circle markers) and the results from \eqref{eq:frenning:M_v2} (solid, dashed and dash dotted line). 
First, we observe that the molecule release process takes several days compared to the exemplary scenario in Fig.~\ref{fig:profile:m}.  
This is due to the large $\nacs$ values and small $\Dm$ values inside the core of the matrix. 
Furthermore, we observe a pH dependence of the DOX release, i.e., the release process takes longer for larger pH values (see \cite{Sutton2007} for a detailed discussion). 
Moreover, the results from the simplified release model \eqref{eq:frenning:M_v2} for DOX and $\beta$-lap are in excellent agreement with the PBS results. 
%\textcolor{blue}{In practice, the release of DOX becomes slower for $t > 100\,\si{\hour}$ than can be seen in Fig.~\ref{fig:real:m}. However, as we neglected long term interactions between DOX molecules and the polymer inside the micelle, the results in Fig.~\ref{fig:real:m} are a good approximation for the actual release of DOX (cf. \cite[Fig.~4]{Sutton2007}).} 

In Fig.~\ref{fig:real:cr}, we investigate the CR of a practical DD system with diblock copolymer micelle matrix TX. 
The figure shows results for PBS (circle markers), the CR \eqref{eq:cr:conv1} (solid lines), and the proposed approximation for $\ratio \gg 1$ in \eqref{eq:cr:beta+:3} (triangle markers).
We assumed that the drug molecules are released from the TX at a distance of $d = 10\,\si{\micro\meter}$ from a cell nucleus or a mammalian cell of size $\rrx = 5\,\si{\micro\meter}$ \cite{SUN2000184}. 
The value for $\Dc$ of DOX has been chosen based on the estimated diffusion coefficient of DOX in non-ablated tumor structures \cite{Weinberg2007}\footnote{For $\beta$-lap, we could not find an exact value for $\Dc$ in the literature, but because of the different values for $\Dm$ of $\beta$-lap and DOX, we used a larger diffusion coefficient for $\beta$-lap.}. 
First, we observe from Fig.~\ref{fig:real:cr} that the duration of the CR is in the order of hours, due to the long duration of the release process $\trel$. 
The ratios $\ratio$ in \eqref{eq:ratio:2} for the different drug molecules are given by $\ratio_{\beta-\mathrm{lap}} \approx 8\cdot 10^{3}$ for $\beta$-lap, and $\ratio_{\mathrm{DOX-pH}\,5.0} \approx 1\cdot 10^{3}$ and $\ratio_{\mathrm{DOX-pH}\,7.4} \approx 1\cdot 10^{4}$ for DOX. 
Therefore, in all three considered cases, the approximation for the \textit{release dominated regime} in \eqref{eq:cr:beta+:3} can be applied and is in excellent agreement with the actual CR \eqref{eq:cr:conv1} in Fig.~\ref{fig:real:cr}.  
%Moreover, we observe from Fig.~\ref{fig:real:cr} that the actual CR and the approximation for $\ratio \gg 1$ are in excellent agreement. 
These results reveal the necessity of employing accurate TX models as the influence of the channel is almost negligible for the considered practical drug carrier and drug molecules when released in the vicinity of the target site. 

% ----------------------------------
% ----------------------------------

%\begin{itemize}
%	\item Describe considered scenario, together with a new picture (TX, RX, Channel, etc.) 
%	\item Describe simulation scenario 
%	\item Describe PBS, especially how the moving boundary is included (mention reflection at undissolved Core) 
%	\item Mention comparison to \cite{noel:16}
%	\item I suggest to use distances 2,5,10, 20 $\rightarrow$ Fig. 3 + another figure. 
%	\item Should we vary anything else in the setting? 
%	\item I suggest to show the influence on the absorption rate over time (relate to applications) 
%	\item Should we show the deviation from volume release? 
%\end{itemize}
%
%\textcolor{red}{From Here -- OLD STUFF}

%% file: conclusions.tex
In this paper, we have modeled a practical polymer-based drug carrier by a spherical homogeneous matrix system. 
We discussed the gradual molecule release from the matrix, which is based on a moving boundary separating dissolved and undissolved molecules. 
We derived expressions for the CR of a matrix TX for an absorbing RX in free space. 
Furthermore, we derived an analytical expression for the CR for the special case of an instantaneous release from the matrix, which is equivalent to the CR of the well-known transparent spherical~TX. 
We proposed a criterion for the characterization of the molecule transport in a DD system and defined two limiting regimes, where either the release process or the channel dynamics determine the characteristic of the CR.
While the exact CR of a matrix TX can usually only be evaluated numerically, we proposed closed-form approximations for the actual CR in both limiting regimes. 
Furthermore, we have shown that the proposed approximations are of particular interest for the characterization of practical DD systems.
%For both limiting regimes, we proposed approximations for the actual CR that can be derived in closed form, while the actual CR can usually only be evaluated numerically.  
Our numerical evaluations showed that the CR of a matrix TX is significantly different from that of a point TX and a transparent spherical TX. 
In particular, matrix TXs spread the release and therefore also the absorption of the molecules over time. 
All presented results have been validated by particle-based simulations. 

Our numerical evaluation revealed the necessity of taking practical drug carrier models into account for the design of controlled-release delivery systems. 
Therefore, it is an interesting topic for future work to further improve the considered matrix TX models by taking into account additional effects such as swelling, erosion, and the reaction of the drug molecules with the polymer matrix structure.
Also, studying the impact of a porous matrix structure and reflections of molecules at the outer boundary of the matrix TX are of interest.
Moreover, investigating the information exchange between nanoparticles, where each particle is modeled as a matrix TX that is loaded with signaling molecules instead of drugs, is a promising research direction \cite{Diez_17}.

%% file: ap_ub_chDom.tex
We define the error for the approximate CR in the \textit{channel dominated regime} as follows
\begin{align}
	\Delta_{\mathrm{CR},\ratio \ll 1}(t) = \tilde{N}(t)\big\vert_{\ratio \ll 1} - N(t) = \Delta_\mathrm{M}(t) \ast p_\mathrm{s}(t),
	\label{eq:cr:beta-:app:1}
\end{align}
where $\tilde{N}(t)\big\vert_{\ratio \ll 1}$ is the approximate CR from \eqref{eq:cr:beta-} and $N(t)$ is the actual CR in \eqref{eq:cr:conv1}.
Furthermore, the error for the release process $\Delta_\mathrm{M}(t)$ is defined as the difference between the actual release process $\nicefrac{\bar{M}(t)}{M_\infty}$ and the instantaneous release as follows
\begin{align}
	\Delta_\mathrm{M}(t) = \epsilon(t) - \frac{\bar{M}(t)}{M_\infty} = 
	\begin{cases}
		1 - \frac{\bar{M}(t)}{M_\infty} & 0\le t \le \trel\\[-2ex]
		0 & t > \trel
	\end{cases}.
	\label{eq:cr:beta-:app:2}
\end{align}
Now, we define an upper bound $\bar{\Delta}_\mathrm{M}$ for the error of the release process $\Delta_\mathrm{M}$ in \eqref{eq:cr:beta-:app:2} as follows
\begin{align}
	\bar{\Delta}_\mathrm{M}(t) = \epsilon(t) - \epsilon(t-\trel) = 
	\begin{cases}
		1  & 0\le t \le \trel\\[-2ex]
		0 & t > \trel
	\end{cases}.
	\label{eq:cr:beta-:app:3}
\end{align}
Although this bound is rather conservative, it facilitates an approximation of the error $\Delta_\mathrm{CR}$ in \eqref{eq:cr:beta-:app:1} in terms of an upper bound $\bar{\Delta}_{\mathrm{CR},\ratio \ll 1}$ as follows
\begin{align}
	\Delta_{\mathrm{CR},\ratio \ll 1}(t) \!=\! \Delta_\mathrm{M}(t) \!\ast\! p_\mathrm{s}(t) \!<\! \bar{\Delta}_{\mathrm{CR},\ratio \ll 1}(t) \!=\! \bar{\Delta}_\mathrm{M}(t)\!\ast\! p_\mathrm{s}(t) &= \!\!\int_0^t \!\!\bar{\Delta}_\mathrm{M}(t)p_\mathrm{s}(t - \xi)\dint{\xi}\nonumber \\ 
	&= \begin{cases}
		\int_0^t p_\mathrm{s}(\xi)\dint{\xi} & \!\!0\le t \le \trel\\[-1ex]
		\int_{t-\trel}^t p_\mathrm{s}(\xi)\dint{\xi} & \!\!t > \trel
	\end{cases}.
	\label{eq:cr:beta-:app:4}
\end{align}
In the \textit{channel dominated regime} $\ratio \ll 1$ holds, and hence, according to \eqref{eq:ratio:2} we have $0 < \trel \ll \tabs$. 
Hence, only the case $t > \trel$ is of interest in \eqref{eq:cr:beta-:app:4}.
As $p_\mathrm{s}(t)$ is almost constant in the integration interval $t-\trel < \xi < t$, the approximation error becomes
\begin{align}
	\Delta_{\mathrm{CR},\ratio \ll 1}(t) < \bar{\Delta}_{\mathrm{CR},\ratio \ll 1}(t) = p_\mathrm{s}(t)\trel.
	\label{eq:cr:beta-:app:5}
\end{align}
Thus, for small values of $\trel$, $\Delta_{\mathrm{CR},\ratio \ll 1}$  becomes small. 
This confirms that the proposed approximation \eqref{eq:cr:beta-} of the CR for the \textit{channel dominated regime} is accurate when the channel dynamic determines the characteristic of the CR.
%is the dominating process, while the exact dynamics of the release process can be omitted. 
%We note, that for $\trel\to 0$, $\ratio \to 0$ and the proposed approximation is equivalent with an instantaneous release of $M_\infty$ particles from the TX.
%A numerical evaluation of the proposed approximation and the error bound is provided in Section~\ref{subsec:validRegimes}.

%% file: ap_ub_reDom.tex
%Utilizing the amount of absorbed molecules $N_\mathrm{s}(t)$ in \eqref{eq:TrSurfTX_N}, an alternative expression for \eqref{eq:cr:conv1} can be obtained by the convolution with the drug release rate $m(t)$ as follows
%\begin{align}
%	N(t) = m(t)\ast N_\mathrm{s}(t) = \int_0^t m(t-\xi)N_\mathrm{s}(\xi)\dint{\xi}, \label{eq:cr:conv2}
%\end{align}
%where $m(t) = \frac{\partial}{\partial t} M(t)$ is the molecule release rate of the matrix TX. 
%Both expressions \eqref{eq:cr:conv1} and \eqref{eq:cr:conv2} are equivalent, as for the convolution holds $\frac{\partial}{\partial t}x(t)\ast y(t) = \frac{\partial}{\partial t}y(t)\ast x(t)$.
%
The error for the approximate CR in the \textit{release dominated regime} can be defined as follows
\begin{align}
	\Delta_{\mathrm{CR},\ratio \gg 1} = \tilde{N}(t)\big\vert_{\ratio \gg 1} - N(t) = m(t) \ast \Delta_{N_\mathrm{s}}(t),
	\label{eq:cr:beta+:app:1}
\end{align}
where $\tilde{N}(t)\big\vert_{\ratio \gg 1}$ is the approximate CR from \eqref{eq:cr:beta+:3} and $N(t)$ is the actual CR. 
%derived from the convolution in \eqref{eq:cr:conv2}. 
In particular, we exploit that the CR can also be derived by the convolution of the release rate $m(t) = \frac{\mathrm{d}}{\mathrm{d} t}M(t)$ and the amount of absorbed molecules $N_\mathrm{s}(t)$ on the right hand side of \eqref{eq:cr:conv1}. 
The error of the amount of absorbed molecules $\Delta_{N_\mathrm{s}}$ is defined as the difference between $N_\mathrm{s}(t)$ and the instantaneous arrival of all particles 
\begin{align}
	\Delta_{N_\mathrm{s}}(t) = \frac{\rrx}{d}\epsilon(t) - N_\mathrm{s}(t) = 
	\begin{cases}
		\frac{\rrx}{d} - N_\mathrm{s}(t) & 0 < t < \tabs\\[-2ex]
		0 & t > \tabs
	\end{cases}.
	\label{eq:cr:beta+:app:2}
\end{align}
Now, we define an upper bound $\bar{\Delta}_{N_\mathrm{s}}$ for $\Delta_{N_\mathrm{s}}$ as follows 
\begin{align}
	\bar{\Delta}_{N_\mathrm{s}}(t) = \frac{\rrx}{d}\left(\epsilon(t) - \epsilon(t - \tabs)\right) = 
	\begin{cases}
		\frac{\rrx}{d} & 0 \leq t \leq \tabs\\[-2ex]
		0 & t > \tabs
	\end{cases}.
	\label{eq:cr:beta+:app:3}
\end{align}
This error bound facilitates an approximation for $\Delta_{\mathrm{CR},\ratio \gg 1}$ in \eqref{eq:cr:beta+:app:1} in terms of an upper bound $\bar{\Delta}_{\mathrm{CR},\ratio \gg 1}$ as follows
\begin{align}
	\Delta_{\mathrm{CR},\ratio \gg 1} \!=\! m(t) \!\ast\! \Delta_{N_\mathrm{s}}(t) \!<\! \bar{\Delta}_{\mathrm{CR},\ratio \gg 1} \!=\! m(t) \!\ast\! \bar{\Delta}_{N_\mathrm{s}}(t) \!&=\!\! \int_0^t \!\!\!m(t - \xi) \bar{\Delta}_{N_\mathrm{s}}(\xi)\dint{\xi}\nonumber\\
	&= \begin{cases}
		\frac{\rrx}{d}\int_0^tm(\xi)\dint{\xi} & \!\!\!0 < t < \tabs \\[-1ex]
		\frac{\rrx}{d}\int_{t - \tabs}^tm(\xi)\dint{\xi} & \!\!\!t > \tabs
	\end{cases}\!.
	\label{eq:cr:beta+:app:4}
\end{align}
In the \textit{release dominated regime} $\ratio \gg 1$ holds, and hence, according to \eqref{eq:ratio:2} we have $0 < \tabs \ll \trel$. 
Hence, only the case $\tabs < t$ is of interest in \eqref{eq:cr:beta+:app:4}.
As $m$ is almost constant in the integration interval $t -\tabs < \xi < t$, the approximation error becomes
\begin{align}
	\Delta_{\mathrm{CR},\ratio \gg 1} < \bar{\Delta}_{\mathrm{CR},\ratio \gg 1} = \frac{\rrx}{d}m(t)\tabs.
	\label{eq:cr:beta+:app:5}
\end{align}
Thus, for small values $\tabs$, $\Delta_{\mathrm{CR},\ratio \gg 1}$ becomes small. This confirms that the proposed approximation \eqref{eq:cr:beta+:3} of the CR in the \textit{release dominated regime} is accurate when the release process mainly determines the characteristic of the CR.

%suitable to be applied when the molecule release process is more dominant in the CR, while the exact dynamics of the channel can be omitted.
%A numerical evaluation of the proposed approximation and the error bound is provided in Section~\ref{subsec:validRegimes}.